\documentclass[draftcls,onecolumn,12pt]{IEEEtran}

\usepackage{cite}
\usepackage{bm}
\usepackage{url}
\usepackage[usenames,dvipsnames]{color}
\usepackage[dvipsnames]{xcolor}
\usepackage{amsmath}
\usepackage{mathtools}
\usepackage[]{amssymb}
\usepackage{amsthm}
\usepackage{float}
\usepackage{dsfont}
\usepackage{array}
\usepackage{booktabs}
\usepackage{multirow}

\usepackage{footmisc}

\usepackage[dvips]{graphicx}
\usepackage[para,flushleft]{threeparttable} 
\usepackage{multirow}	
\usepackage{dcolumn} 

\usepackage[ruled]{algorithm2e}
\usepackage{xifthen} 

\makeatletter
\newcommand{\removelatexerror}{\let\@latex@error\@gobble}
\makeatother

\newtheorem{theorem}{Theorem}
\newtheorem{lemma}[theorem]{Lemma}
\newtheorem{remark}{Remark}

\DeclareMathOperator{\diag}{diag}
\DeclareMathOperator{\vect}{vec}

\DeclareMathOperator{\KL}{KL}

\newif \ifwebcolor
\webcolortrue

\def \ba {\begin{array}}
\def \ea {\end{array}}
\def \benu {\begin{enumerate}}
\def \eenu {\end{enumerate}}
\def \bdes {\begin{description}}
\def \edes {\end{description}}

\def \bitem {\begin{itemize}}
\def \eitem {\end{itemize}}

\def \bfl {\begin{flushleft}}
\def \efl {\end{flushleft}}
\def \bfr {\begin{flushright}}
\def \efr {\end{flushright}}

\def \beq {\begin{equation}}
\def \eeq {\end{equation}}
\def \bqa {\begin{eqnarray}}
\def \eqa {\end{eqnarray}}
\def \bqa* {\begin{eqnarray*}}
\def \eqa* {\end{eqnarray*}}

\def \bal {\begin{align}}
\def \eal {\end{align}}

\renewcommand{\baselinestretch}{1.7}
\newcounter{mytempeqncnt}

\DeclareMathOperator*{\argmax}{arg\,max}
\DeclareMathOperator*{\argmin}{arg\,min}

\begin{document}
%
\title{Semi-blind Channel Estimation and Data Detection for Multi-cell Massive MIMO Systems on Time-Varying Channels}

\author{\IEEEauthorblockN{Mort Naraghi-Pour, ~\IEEEmembership{Senior Member,~IEEE}, Mohammed Rashid, ~\IEEEmembership{Student Member,~IEEE},
Cesar Vargas-Rosales, ~\IEEEmembership{Senior Member,~IEEE}}
\thanks{
Mort Naraghi-Pour and Mohammed Rashid are with the Division of Electrical and Computer Engineering, 
School of Electrical Engineering and Computer Science, Louisiana State University, Baton Rouge, LA 70803
\{email: naraghi@lsu.edu, mrashi2@lsu.edu\}.
Cesar Vargas-Rosales is with Tecnologico de Monterrey, 
Monterrey, Nuevo León, México \{email:cvargas@tec.mx\}}.}


\maketitle
\thispagestyle{empty}
\pagestyle{empty}

\def\bda{\mathbf{a}}
\def\bdg{\mathbf{g}} 
\def\bdh{\mathbf{h}}
\def\bdm{\mathbf{m}}
\def\bds{\mathbf{s}} 
\def\bdv{\mathbf{v}}
\def\bdw{\mathbf{w}} 
\def\bdx{\mathbf{x}} 
\def\bdy{\mathbf{y}} 
\def\bdz{\mathbf{z}}
\def\bdA{\mathbf{A}}
\def\bdC{\mathbf{C}}
\def\bdD{\mathbf{D}} 
\def\bdF{\mathbf{F}}
\def\bdG{\mathbf{G}} 
\def\bdH{\mathbf{H}}
\def\bdI{\mathbf{I}}
\def\bdJ{\mathbf{J}}
\def\bdK{\mathbf{K}}
\def\bdQ{\mathbf{Q}}
\def\bdR{\mathbf{R}}
\def\bdS{\mathbf{S}}
\def\bdV{\mathbf{V}}
\def\bdX{\mathbf{X}}
\def\bdGamma{\bm{\Gamma}}
\def\bdmu{\bm{\mu}}
\def\bdSigma{\bm{\Sigma}}
\def\bdzeta{\bm{\zeta}}
\def\bdLambda{\bm{\Lambda}}
\def\bdeta{\bm{\eta}}

\def\setA{\mathcal{A}} 
\def\bdsetS{\bm{\mathcal S}}
\def\bdsetH{\bm{\mathcal H}}
\def\bdsetY{\bm{\mathcal Y}}
\def\famF{\mathcal{F}}
\def\tq{\tilde{q}}
\def\tbdJ{\tilde \bdJ}
\def\l{\ell}
\def\bdzero{\mathbf{0}} 
\def\Exp{\mathbb{E}} 
\def\M{\mathcal{M}} 
\def\R{\mathbb{R}} 
\def\C{\mathbb{C}} 
\def\CN{\mathcal{CN}} 
\def\N{\mathcal{N}} 
\def\Es{E_s} 

\def\symset{\setA_\M} 
\def\symsetK{\symset^K} 
\begin{abstract}
We study the problem of semi-blind channel estimation and 
symbol detection in the uplink of
multi-cell massive MIMO systems with spatially correlated time-varying channels. 
An algorithm based on expectation propagation (EP) is developed to iteratively
approximate the joint a posteriori distribution of the unknown
channel matrix and the transmitted data symbols with a
distribution from an exponential family. This distribution is
then used for direct estimation of the channel matrix and
detection of data symbols.
A modified version of the popular Kalman filtering 
algorithm referred to as KF-M emerges from our EP derivation
and it is used to initialize the EP-based algorithm. Performance of the 
Kalman smoothing algorithm 
followed by KF-M is also examined. Simulation
results demonstrate 
that channel estimation error and the symbol error rate (SER)
of the semi-blind KF-M, KS-M, and EP-based algorithms improve 
with the increase in the number of base station antennas and the length of the data symbols 
in the transmitted frame. 
It is shown that the EP-based algorithm significantly outperforms KF-M
and KS-M algorithms in channel estimation and symbol detection.
Finally, our results show that when applied to time-varying channels, 
these algorithms outperform 
the algorithms that are developed for block-fading channel models.

\end{abstract}


\section{Introduction}

In wireless communication, coherent demodulation of transmitted symbols requires accurate knowledge of 
channel state information (CSI).
Traditionally, pilot sequences are transmitted to aid in channel estimation. 
Since the number of transmit antennas determines the length of the 
orthogonal pilot sequences,
for massive MIMO systems employing a large number of antennas at the base station (BS),
pilot-based channel estimation in the 
downlink is very challenging.
Due to the increased length of pilot sequences,
the time required for pilot transmission increases resulting in low spectral efficiency.
In addition, the time required for data and pilot transmission may exceed the 
coherence time of the channel.

In time division duplexing (TDD), 
CSI can be estimated in the uplink from pilot transmissions. 
For users employing a single antenna, the length of the pilot sequence 
only needs to scale with the number of users
in the cell which is typically much smaller than the number of BS antennas.  
Invoking the channel reciprocity property, the uplink CSI matrix is 
transposed to obtain the downlink CSI which
can then be used for precoding in the downlink. 
However, even in this case, due to the limited number of orthogonal 
pilot sequences, pilots must be shared among the users 
in the neighboring cells resulting in the so-called pilot contamination
problem \cite{Elijah}, which diminishes the accuracy of CSI estimation. 

To overcome the effects of pilot contamination, several
blind channel estimation schemes have been proposed in recent years
\cite{Ghavami3, zhang, Yuan, Mezh}.
In \cite{Ghavami3} the expectation propagation (EP)
algorithm \cite{minka2001expectation} is developed for blind 
channel estimation and symbol detection in multi-cell massive MIMO systems.
Bilinear generalized approximate message passing (BiG-AMP) 
algorithm is another approach proposed in \cite{zhang} 
for sparse massive MIMO channels. 
Channel sparsity is exploited in a number of other studies including
\cite{Yuan, Mezh, Liu_Blind}. In particular
\cite{Mezh} studies the effect of one-bit ADCs in the receiver.

Blind channel estimation methods suffer from inherent 
phase ambiguities in the demodulated symbols, 
requiring a pilot symbol and user label 
in order to resolve these ambiguities \cite{zhang}. 
In contrast, adding a few additional training symbols in a semi-blind 
approach significantly improves 
the channel estimation accuracy compared to the blind methods (see Fig. $1$ in \cite{Carvalho_bounds, SB_algo2}). 
A semi-blind channel estimation method based on the expectation maximization algorithm is proposed 
and analyzed in \cite{Nayebi}.
A semi-blind pilot decontamination scheme is proposed for
a multi-cell massive MIMO system  in \cite{DHu} 
where the information sequence of the target cell is first 
estimated and used as a pilot sequence in a least square method for CSI estimation. 
In \cite{Liang_Semiblind} the authors present a semi-blind 
joint channel estimation and data detection method based on the
regularized alternating least-square (R-ALS) method. 
Other semi-blind algorithms for 
massive MIMO channels are proposed in 
\cite{SB_algo1,SB_algo2} where the pilots sent in the 
beginning portion of the frames are used for their 
initialization. For sparse massive MIMO channels, a message passing semi-blind channel estimation algorithm is proposed in \cite{Yan}. 

In the semi-blind algorithms described above, the massive MIMO channel 
is assumed to be spatially uncorrelated, as well as 
static during the transmission time of a frame. 
Due to their large number of antennas,
the BS's in massive MIMO systems have fine angular resolutions
making some spatial directions more probable than others 
\cite{Li-wang, Yang-qin, Bjornson}. 
This results in spatial correlation in the channel vector of a user 
to the BS which needs to be included in the channel model. 
Secondly, while the assumption of static channel 
(the so-called block-fading) model
is valid for stationary or low-mobility users, it breaks down for 
high-mobility users.
In addition to delay spread due to multipath effects, 
mobile fading channels are subject to time variations due to Doppler spread.
For block-fading channels CSI estimation is only required at the start of each frame.
However, for time-varying channels, the CSI estimates need to be updated instantaneously throughout the frame. 
Compared to the rich literature available on block-fading 
channel estimation in massive MIMO systems,
few studies are available for the estimation of the time-varying channels.
For a time-varying massive MIMO channel, the data rates achievable by the linear MMSE channel 
estimator is studied in \cite{Rates_TV,MMSE_TV, Papa_CA, AR_MassiveMIMO}. 
The time evolution of the channel taps is modeled by a first-order autoregressive (AR) process with the temporal 
correlation properties corresponding to the Jakes' model \cite{Jakes_book}.
Kalman filter is used to estimate the time-varying sparse massive MIMO 
channel in \cite{Aditya_SBL} and the 
time-varying non-sparse massive MIMO channel in 
\cite{KF_TDD_TV, KF_mMIMO}.

In this paper, we consider
semi-blind joint channel estimation and data detection in multi-cell 
massive MIMO systems for a spatially correlated 
time-varying channel. To our knowledge, this problem has not been studied for massive 
MIMO systems. The contributions made in this paper are summarized as follow:

\begin{itemize}
\item We consider a multi-user multi-cell massive MIMO system.
The uplink channel from each user to BS is assumed to be spatially correlated
and time-varying. The channel vector from a user to the BS is modeled 
as a complex circularly-symmetric Gaussian vector with a given 
(known) correlation matrix \cite{Bjornson}.
The temporal correlation of the channel is modeled by a first-order 
AR process, \cite{Rates_TV,MMSE_TV,AR_MassiveMIMO}
with correlation properties corresponding to the Jakes' model \cite{Jakes_book}. 
A semi-blind method is developed 
for joint channel estimation and data detection where
it is assumed that the users transmit a few pilot symbols
(on the order of the number of users) at the beginning of each
frame. These symbols are used for an initial estimation of the channel
and to overcome the inherent ambiguity of non-coherent
detectors.
The proposed method is based on the
expectation propagation (EP) algorithm 
and iteratively approximates the joint a posterior distribution of the
channel matrix and the transmitted data symbols with a distribution from 
an exponential family. 
This distribution is then used for direct estimation of the channel matrix and detection of the data symbols. 

\item Simulation results are presented to demonstrate the performance of the 
proposed EP-based algorithm (EP) in terms of channel estimation and symbol error rate (SER).

A modified version of the popular Kalman filtering (KF) algorithm 
referred to as KF-M is also proposed. KF-M emerges from our EP derivation 
and is used to initialize the EP-based algorithm.
The backward recursion equations of Kalman smoother are also a part of our 
EP derivations.
Therefore, the performance of KF-M as well as
the Kalman smoothing algorithm followed
by a single pass of KF-M is also presented here for comparison and
is denoted as KS-M. To benchmark the performance of semi-blind
KF-M, KS-M, and EP, we also present the performance
of the Kalman filter and smoother in a pure training mode
(TM) when the entire frame is composed of known pilot
symbols and only channel estimation is performed. These
two cases, are referred to as KF-TM and KS-TM. In channel
estimation, KF-TM provides a lower bound for KF-M, and
KS-TM provides a lower bound for KS-M, and EP. Finally,
we also plot the SER performance of the MMSE estimator
with known CSI (denoted PCSI) for comparison with SER
performance of the proposed algorithms.

\item To our knowledge the problem under consideration here has not been previously
investigated. As such, unfortunately
we cannot compare our results with those in  published literature.
However, to verify that algorithms developed under the assumption
of a block-fading channel
are not suitable for a time-varying channel model, 
we compare our results with those from \cite{Nayebi} and \cite{Liang_Semiblind}. 
The comparison shows that for time-varying channels,
the proposed method significantly outperforms the methods
in \cite{Nayebi} and \cite{Liang_Semiblind}.
\end{itemize}

The rest of this paper is organized as follows. Section \ref{sys-model} 
describes the system model of a time-varying multicell massive MIMO system. 
The semi-blind EP algorithm for this system is derived in section \ref{algorithm}. 
Simulation results are discussed in section \ref{s-results} and 
section \ref{conclusn} concludes this work.
   
%
\textit{{Notations}:} Throughout this paper, small letters $(x)$ are used for 
scalars, bold small letters $(\bdx)$ for vectors, and bold capital letters 
$(\bdX)$ for matrices. $\R$ and $\C$ represent the set of real and complex 
numbers, respectively. 
The superscripts $(.)^T$, $(.)^H$, $(.)^*$, and $(.)^{-1}$ represent transpose, 
Hermitian transpose, complex conjugate, and matrix inverse, respectively. Also, $\otimes$ denotes 
the matrix Kronecker product. 
For a pdf $p(.)$, 
$\Exp_{p}$ denotes the expectation operator with respect to $p(.)$.
$\bdI_N$ denotes the $N \times N$ identity matrix.
Finally, $\vect(\bdX)$ and $||\bdx||$ denote the 
vectorization of the matrix $\bdX$ and the $\l^2$ norm of the vector 
$\bdx$, respectively.
\section{System Model} \label {sys-model}

We consider a multi-user MIMO network made up of $L$ cells each with its own BS and with $K$ users located inside every cell. 
Every BS in a cell has $M$ antennas and each user has a single-antenna transceiver.  
At time $t$, the channel gain between the $m$-th antenna of the $l$-th BS and the $k$-th 
user present in the $i$-th cell is represented by ${h}_{limk}(t)$. Each channel 
gain ${h}_{limk}(t)$ can be written as 
\begin{equation} \label{eq_1}
{h}_{limk}(t) = {g}_{limk}(t) \sqrt{\beta_{lik}},
\end{equation}
where, ${g}_{limk}(t)$ models the fast-fading channel between the $k$-th 
user in cell $i$ and the $m$-th antenna of BS $l$, and $\beta_{lik}$ models 
the large-scale fading incurred by the geometric attenuation and shadowing effects. 
We assume that $\beta_{lik}$ is a known constant which is independent of the antenna index $m$.

The fast-fading channel ${g}_{limk}(t)$ is
considered to be a wide-sense stationary complex Gaussian process with
zero mean and unit power. 
Using the Jakes' model \cite{Jakes_book}, the time autocorrelation of ${g}_{limk}(t)$ 
is given by
\begin{align}\label{R_g_eq}
R^g_{limk}\left(\Delta t\right)&=\Exp \left[{g}_{limk}(t){g}^{*}_{limk}(t+\Delta t)\right]\nonumber \\
&=J_0\left(2\pi f^d_{limk}|\Delta t|\right),
\end{align}
in which, $J_0(.)$ is the zero-order Bessel function of first kind and $f^d_{limk}$ represents the normalized maximum Doppler shift corresponding to the channel between the $m$-th antenna of cell $l$ and the $k$-th user in cell $i$. 
The time autocorrelation 
function of $h_{limk}(t)$ can be obtained as
\begin{align}\label{R_h_eq}
R^h_{limk}\left(\Delta t\right)&=\Exp \left[{h}_{limk}(t){h}^{*}_{limk}(t+\Delta t)\right]\nonumber \\
&=\beta_{lik}\Exp \left[{g}_{limk}(t){g}^{*}_{limk}(t+\Delta t)\right]\nonumber \\
&=\beta_{lik}J_0\left(2\pi f^d_{limk}|\Delta t|\right),
\end{align}
Let ${\bdg_{lik}(t)} \triangleq [{g}_{li1k}(t), 
{g}_{li2k}(t), \ldots, {g}_{liMk}(t)]^T$ represent the $M \times 1$ 
fast-fading channel vector from the 
$k$-th user in cell $i$ to the 
$l$-th BS antenna array. 
We assume that the elements in $\bdg_{lik}(t)$ are correlated
with the correlation matrix
$\bdR_{lik} \triangleq \Exp[\bdg_{lik}(t)\bdg^H_{lik}(t)]$
\cite{Forenza_SC,Kronecker_SC}. 
Let ${\bdG}_{li}(t) \triangleq
[{\bdg}_{li1}(t), \ldots, {\bdg}_{liK}(t)]$. The overall channel 
gain between the $l$-th BS and the users in cell $i$ is given by
\begin{equation}
\bdH_{li}(t) \triangleq {\bdG}_{li}(t)\bdD_{li}^{\frac{1}{2}},
\end{equation} 
where $\bdD_{li} \triangleq \diag\{\beta_{li1},\beta_{li2}, \ldots, 
\beta_{liK}\}$. 
It is assumed that the users' channels to the BS are
uncorrelated. In particular this implies that 
$\Exp[{\bdG}_{li}(t){\bdG}_{lj}^H(t)] = \sum^K_{k=1} \Exp[\bdg_{lik}(t)\bdg^H_{ljk}(t)]=\sum^K_{k=1} \bdR_{lik} \delta_{ij}$, and as a result,
\begin{align}
\Exp[\bdH_{li}(t) \bdH_{lj}^H(t)]
 &= \Exp[\bdG_{li}(t) \bdD_{li} \bdG_{li}^H(t)]\delta_{ij},
\nonumber\\
& = \sum_{k=1}^K \beta_{lik}\bdR_{lik} \delta_{ij},
\end{align}

The signal vector received at BS $l$ at time $t$ is
given by
\begin{align}
\bdy_l(t) &= \sum_{i=1}^{L} \bdH_{li}(t) \bds_{i}(t) + \bdw'_l(t) \nonumber \\
          &= \underbrace{\bdH_{ll}(t) \bds_{l}(t)}_{\text{desired signal}} 
             + \underbrace{\sum_{\stackrel{i=1}{i\neq l}}^{L} \bdH_{li}(t) \bds_{i}(t)}_{\text{interference}} + \underbrace{\bdw'_l(t)}_{\text{noise}},
\label{eq_model} 
\end{align}
where $\bds_{i}(t) = [s_{i1}(t) ,s_{i2}(t), \ldots, s_{iK}(t)]^T$ represents the 
transmitted symbols by all the users in 
the $i$-th cell. We assume that the symbols $s_{ij}(t)$ 
belong to an $\M$-ary modulation constellation, denoted by $\symset$, 
and have zero mean with average energy $\Es$.
We also assume that $\Exp[\bds_{i}(t) \bds_{j}(t)^H] = \Es \delta_{ij} \bdI_K$, i.e., the symbols $s_{ij}(t)$ are independently selected from $\symset$. 
The noise term at the $l$-th BS is modeled with a circularly symmetric complex Gaussian distribution, i.e., $\bdw'_l(t) \sim \CN(\bdw'_l|\bdzero, \bdI_M)$. 

Let  $\bdw_l(t) \triangleq 
\sum_{\stackrel{i=1}{i\neq l}}^{L} \bdH_{li}(t) \bds_{i}(t) + \bdw'_l(t)$ 
denote the overall disturbance at the $l$-th BS. 
Thus, \eqref{eq_model} can be written as $\bdy_l(t) = \bdH_{ll}(t) \bds_{l}(t) + 
\bdw_l(t)$ where $\bdw_l(t)$ has zero mean and correlation matrix 
\begin{align}
\bdR_w \triangleq \Exp[\bdw_l(t) \big( \bdw_l(t) \big )^H]
	= \Es \sum_{\stackrel{i=1}{i\neq l}}^{L} \sum_{k=1}^K \beta_{lik}\bdR_{lik} + \bdI_M,\label{eq_newNoisePower}
\end{align}

Assuming that $KL$ is large and using the central limit theorem we consider $\bdw_l(t)$
as a circularly symmetric complex-valued Gaussian vector, i.e., 
$\bdw_l(t) \sim \CN \left( \bdw_l | \bdzero, \bdR_w \right)$. 

To keep the notations uncluttered, in the following we drop the subscript $l$ from 
the signal model and write the received vector at time $t$ at the $l$-th BS as   
\begin{equation}
\bdy_t = \bdH_t \bds_t + \bdw_t,
\label{eq_MIMOmodelComplex}
\end{equation}
where $\bdH_t$ is the overall channel gain matrix, $\bds_t \in \symset^K$ represents 
the transmitted symbols by the $K$ users, and $\bdw_t$ is the zero mean 
complex Gaussian distributed disturbance with covariance matrix given by \eqref{eq_newNoisePower}.

By applying the vectorization property as in \cite{Ghavami3}, \eqref{eq_MIMOmodelComplex} can also be 
written as
\begin{equation}
\bdy_t = \bdS_t \bdh_t + \bdw_t,
\label{eq_VEC}
\end{equation}
where we define $\bdS_t = \bds_t^T \otimes \bdI_M$ and 
$\bdh_t = \vect(\bdH_t)$.
\begin{figure*}[ht]
    \centering
     	\includegraphics[width=0.99\textwidth]{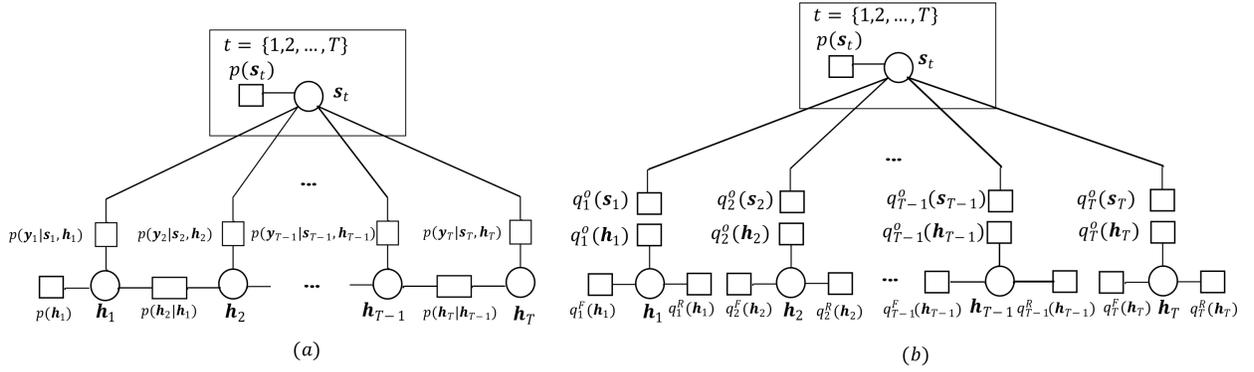}
	\caption{Factor graph illustrations of $(a)$ True posterior distribution in \eqref{eq_simoapp}, and 
	$(b)$ Approximated posterior distribution in \eqref{prod-form}. Small rectangles represent factor nodes and 
	circles represent variable nodes. A plate (big rectangle) notation is used 
	to represent a repetition of variables in the subgraph.}
	\label{fig:True_and_Approx_Posterior}
\end{figure*}

An autoregressive (AR) process has been extensively 
used to model the time evolution of the channel matrix \cite{Badd, Tsat, Kom, Qi}. 
Since any higher-order AR model 
can be written as a first-order model in matrix state-space form \cite{Kom, AR_MassiveMIMO}, in this paper we use a first-order AR model (AR(1)) for the 
time-varying vector channel given by
\begin{align}
\bdh_t = \bdA \bdh_{t-1} + \bdv_t,
\label{AR}
\end{align}
where $\bdA$ is a diagonal matrix with the elements on the diagonal denoted by 
$[\bdA]_{n,n}=a_n$, for $n=1,2,\dots,MK$. The variable $a_n$ is 
the AR(1) coefficient corresponding to the $n$-th channel between a user 
and a BS antenna in the cell. We let 
$a_n=J_0(2\pi f^d_n)$ in which $f^d_n$ is the normalized maximum 
Doppler shift for channel $n$. 
The innovation process $\{\bdv_t\}$ is an iid circularly symmetric complex Gaussian process with 
${\bdv}_t \sim \CN (\bdzero, \bdQ)$. To include the spatial correlation of the 
channel vector, we set 
$\bdQ=\bdR^{1/2}_h\bdQ_v\bdR^{1/2}_h$ 
in which, given the independence of users' channels,
$\bdR_h \triangleq 
\Exp[\bdh_t\bdh^H_t]=\diag\{\beta_1\bdR_1,\beta_2\bdR_2,\ldots,\beta_K\bdR_K\}$ 
is a block diagonal matrix. 
The matrix $\bdQ_v$ is diagonal with elements on the diagonal given by $[\bdQ_v]_{n,n}=\sigma^2_n$, 
for $n=1,2,\dots,MK$. To match the autocorrelation function in \eqref{R_h_eq} such that the average 
power of the channel coefficient in \eqref{eq_1} is equal to the large-scale fading coefficient, we set 
$\sigma^2_n=(1-a^2_n)$.

Consider a transmitted frame of length $T$ which is comprised of $T_p$ pilot symbols in the 
beginning followed by $T_d=T-T_p$ unknown data symbols denoted as $\bdsetS 
\triangleq \left(\bdsetS_p,\bdsetS_d\right)$ where we define 
$\bdsetS_p=\left(\bds_1, \bds_2, \ldots,\bds_{T_p}\right)$ and $\bdsetS_d=\left(\bds_{T_p+1},\ldots, \bds_T\right)$. 
The corresponding received vectors are given by $\bdsetY \triangleq \left(\bdsetY_p,\bdsetY_d\right)$,
$\bdsetY_p=(\bdy_1,\bdy_2,\ldots,\bdy_{T_p})$ and $\bdsetY_d=(\bdy_{T_p+1},\ldots,\bdy_T)$. 
Similarly, the channel vector is denoted by 
$\bdsetH \triangleq \left(\bdsetH_p,\bdsetH_d\right)= (\bdh_1, \bdh_2, \ldots,\bdh_{T_p},\bdh_{T_p+1},\ldots, \bdh_T)$.  
We are interested in a detector where, having received $\bdsetY$, 
the unknown channel vectors in $\bdsetH$ and the unknown transmitted symbols in $\bdsetS_d$ 
are jointly estimated. The posterior joint distribution of $\bdsetS$ and 
$\bdsetH$ is given by
\begin{align}
p(\bdsetS, \bdsetH | \bdsetY) 
  &\propto p(\bdsetS_d,\bdsetH) p(\bdsetY_p | \bdsetS_p, \bdsetH_p) p(\bdsetY_d | \bdsetS_d, \bdsetH_d)
  \nonumber \\
  &= \Big[ \prod_{t=T_p+1}^{T} p(\bds_t) \Big] \Big[ \prod_{t=1}^T p(\bdh_t | \bdh_{t-1})
    p(\bdy_t | \bds_t, \bdh_t) \Big]  \nonumber \\
  &= \prod_{t=1}^{T} p(\bds_t) p(\bdh_t | \bdh_{t-1})
    p(\bdy_t | \bds_t, \bdh_t),
\label{eq_simoapp}
\end{align}
in which $p(\bds_t)$ is the probability mass function (pmf) of the
transmitted vector $\bds_t$ and by convention we set $p(\bds_t)=1$ for $t=1,\ldots,T_p$ 
and $p(\bdh_1)=p(\bdh_1|\bdh_0)$.
From \eqref{eq_VEC} we have
\begin{align} \label{eq_chdist}
p(\bdy_t | \bds_t, \bdh_t) = \CN(\bdy_t | \bdS_t \bdh_t, \bdR_w),
\end{align}

The optimum receiver implements the maximum a posteriori rule according to
\eqref{eq_MAP}, i.e.,
\begin{equation}
\label{eq_MAP}
(\bdsetS_d, \bdsetH)^* = \argmax_{\bdsetS_d \in \setA_{\M}^{KT_d}, \bdsetH \in \C^{M\times T}} 
                         p(\bdsetS, \bdsetH | \bdsetY),
\end{equation}
Due to the complexity of \eqref{eq_MAP}, finding the optimum solution is 
generally very difficult and requires multidimensional integration. The 
proposed EP algorithm in the next section exploits the multiplicative nature 
of \eqref{eq_simoapp} to find a simpler approximation for the conditional 
joint distribution of $(\bdsetS, \bdsetH)$ such that the marginals can be 
calculated with much less effort.  

\section{Semi-blind EP formulation}  \label{algorithm}

In this section we develop the EP algorithm for noncoherent semi-blind detection in 
massive MIMO systems for fast fading channel. For a review of the EP algorithm 
we refer to \cite{MinkaPHD, Ghavami1}.
Let $\famF$ denote a family of exponential distributions. Similar to \cite{Qi} we exploit 
the factorized structure of \eqref{eq_simoapp} to approximate the posterior 
distribution $p(\bdsetS, \bdsetH | \bdsetY)$ with the following distributions from $\famF$. 
\begin{equation}
p(\bdsetS, \bdsetH | \bdsetY) \approx q(\bdsetS, \bdsetH) 
= \prod_{t=1}^{T} q_t(\bds_t, \bdh_t)
= \prod_{t=1}^{T} q_t(\bds_t) q_t(\bdh_t),
\label{app.1}
\end{equation}
where $q_t(\bds_t, \bdh_{t}) = q_t(\bds_t) q_t(\bdh_t)$ $\in \famF$. 
Examining \eqref{eq_simoapp} and following \cite{Qi},
we use the following 
product form for ${q}(\bdsetS, \bdsetH)$: 
\begin{align}
q(\bdsetS, \bdsetH) 
  & \propto 
	p(\bds_1) p(\bdh_{1}) 
            q_1^O(\bds_1, \bdh_{1}) \times \nonumber \\
	& \prod_{t=2}^{T} p(\bds_t) q_t^{FR}(\bdh_{t-1}, \bdh_{t}) 
            q_t^O(\bds_t, \bdh_{t}),
\label{eq_ap_approx}
\end{align}
where, comparing \eqref{eq_ap_approx} and \eqref{eq_simoapp}, we have $q_t^{FR}
(\bdh_{t-1},\bdh_{t})$ to approximate $p(\bdh_{t} | \bdh_{t-1})$ and $q_t^O
(\bds_t, \bdh_{t})$ to approximate $p(\bdy_t | \bds_t, \bdh_t)$. Finally, to 
write \eqref{eq_ap_approx} in a completely factorized form as in \eqref{app.1}
we let
\begin{align}
q_t^O(\bds_t, \bdh_{t}) = q_t^O(\bds_t) q_t^O(\bdh_{t}),
\label{p.1}
\end{align}
and 
\begin{align}
q_t^{FR}(\bdh_{t-1}, \bdh_{t}) = q_t^{R}(\bdh_{t-1})q_t^{F}(\bdh_{t}),
\label{p.2}
\end{align}
Now inserting \eqref{p.1} and \eqref{p.2} into \eqref{eq_ap_approx}, and 
letting $q_1^F(\bdh_{1})$ to approximate $p(\bdh_{1})$ and $q_T^R(\bdh_{T}) = 
1$, we can write
\begin{align}
q(\bdsetS,\bdsetH) \propto 
\Big[ \prod_{t=1}^{T} p(\bds_t) q_t^O(\bds_t) \Big] \Big[ \prod_{t=1}^{T} 
q_t^{F}(\bdh_{t}) q_t^{R}(\bdh_{t}) q_t^O(\bdh_{t}) \Big],
\label{prod-form}
\end{align}
From \eqref{prod-form} and \eqref{app.1}, the approximating posterior 
distribution $q_t(\bds_t, \bdh_t)$ is given by
\begin{align}
q_t (\bds_t, \bdh_t) \propto p(\bds_t) q_t^R(\bdh_{t}) q_t^F(\bdh_{t}) 
q_t^O (\bds_t) q_t^O (\bdh_{t}),
\label{post}
\end{align}
Note that in \eqref{post}, from \eqref{eq_simoapp} we have $p(\bds_t)=1$ for all $t=1,\ldots,T_p$. 
The true posterior distribution in \eqref{eq_simoapp} and the approximate one in \eqref{prod-form} 
are illustrated with factor graphs in Fig. \ref{fig:True_and_Approx_Posterior}. Updating the factors $q^F_t(.)$, $q^R_t(.)$, and 
$q^O_t(.)$ in \eqref{post} result in forward, reverse, and observation messages,
respectively, which propagate 
through the factor graph in Fig. \ref{fig:True_and_Approx_Posterior} (a) until a good approximation to the true posterior 
is obtained.

With EP, we update the factors ensuring that they belong to 
the exponential family $\famF$. As a result, their product $q(\bdsetS,
\bdsetH)$ also belongs to $\famF$ and can be effectively used for maximum a 
posteriori estimation. To this end, we select the distributions $q_t^O
(\bdh_{t})$, $q_t^F(\bdh_{t})$ and $q_t^R(\bdh_{t})$ from the exponential 
family $\famF$ and $q_t^O(\bds_t)$ to be discrete. Since $p(\bds_t)$ is also 
discrete, it follows that \eqref{post} and \eqref{prod-form} are from the 
exponential family. In the following, we update these factors.
\begin{figure}[tp]
        \centering
     	\includegraphics[width=0.36\textwidth]{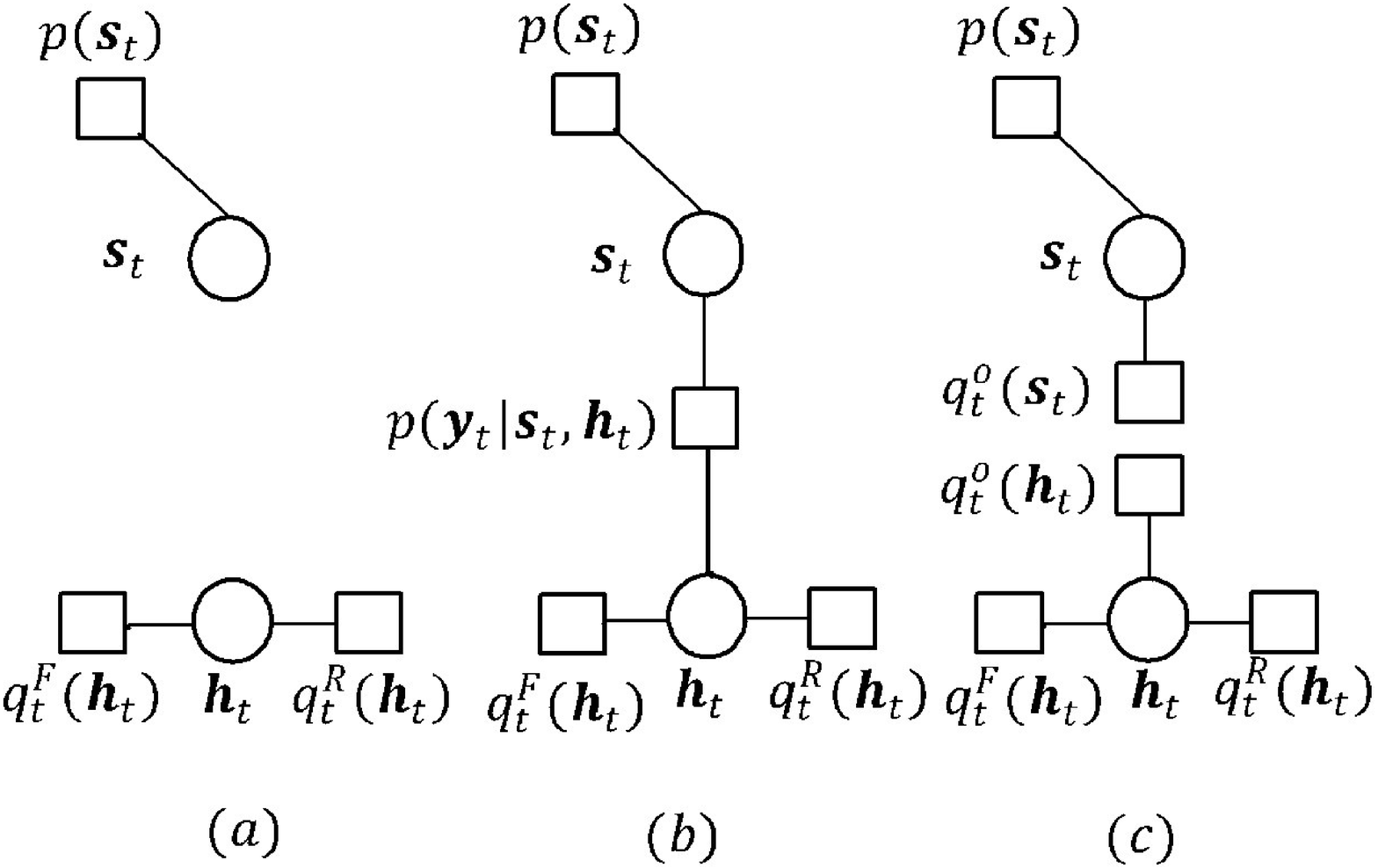}
	\caption{EP steps for updating $q^O_t(\bds_t,\bdh_t)$: $(a)$ Eliminate $q^O_t(\bds_t)$ and $q^O_t(\bdh_t)$ from the factor graph 
to find the cavity distribution $q^{\backslash O}_t(\bds_t,\bdh_t)$ as in \eqref{cavity_0}, $(b)$ 
Use $p(\bdy_t|\bds_t,\bdh_t)$ factor to define the 
hybrid posterior distribution $\hat{q}_t(\bds_t, \bdh_t)$ as in \eqref{eq_int_approx_post}, and 
$(c)$ Project $\hat{q}_t(\bds_t, \bdh_t)$ onto $\famF$ and update 
$q^O_t(\bds_t,\bdh_t)$ as in \eqref{update_q_O_ht} and \eqref{st_est}.}
	\label{fig:Q_O_factor_computation}
\end{figure}
\begin{figure}[tp]
        \centering
     	\includegraphics[width=0.45\textwidth]{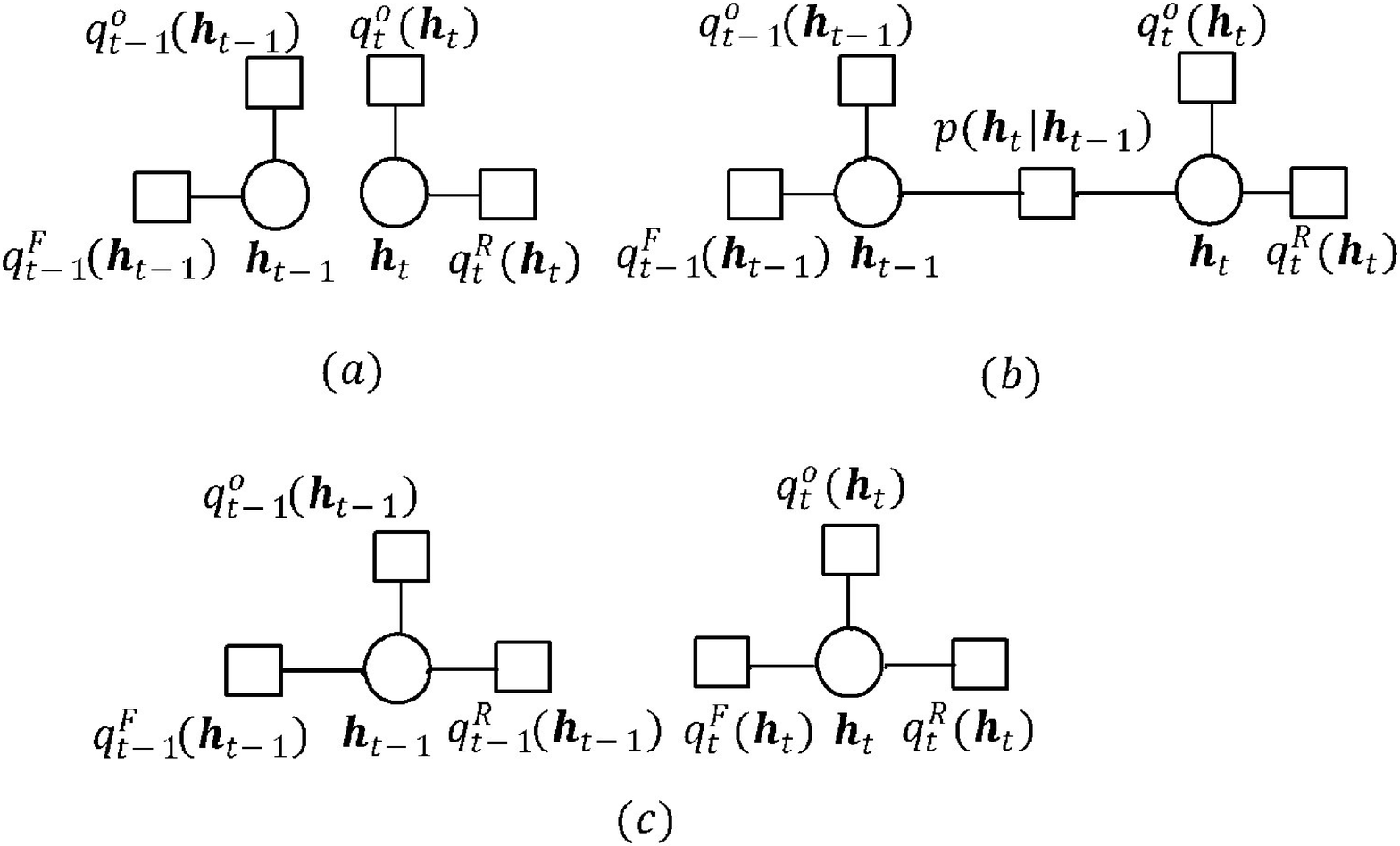}
	\caption{EP steps for updating $q^R_{t-1}(\bdh_{t-1})$ and $q^F_t(\bdh_t)$: $(a)$ 
	Eliminate $q^R_{t-1}(\bdh_{t-1})$ and $q^F_t(\bdh_t)$ from the factor graph to find the 
	cavity distributions $q^{\backslash R}_{t-1}(\bdh_{t-1})$ and $q^{\backslash F}_t(\bdh_t)$ as in \eqref{cavity_FR}, $(b)$ Use $p(\bdh_t|\bdh_{t-1})$ factor to define the intermediate posterior distribution $r_t(\bdh_{t-1},\bdh_t)$ as in 
	\eqref{eq_CNtrans}, and $(c)$ Project $r_t(\bdh_{t-1},\bdh_t)$ onto $\famF$ and update $q^F_t(\bdh_t)$, $q^R_{t-1}(\bdh_{t-1})$ as in \eqref{eq_KLtrans}, \eqref{qF_factor}, and \eqref{qR_factor}.}
	\label{fig:Forward_Reverse_computation}
\end{figure}

First, we ignore the beginning portion of the frame and compute $q_t^O(\bds_t)$ and $q_t^O(\bdh_t)$ 
for the blind part of the frame, 
i.e., for $t=T_p+1,\ldots,T$. As shown in Fig. \ref{fig:Q_O_factor_computation} we define the cavity distribution as
\begin{equation}\label{cavity_0}
q_t^{\backslash O}(\bds_t,\bdh_t)=\frac{q_t(\bds_t,\bdh_t)}{q_t^O(\bds_t) q_t^O(\bdh_t)}=p(\bds_t)q_t^{\backslash O}(\bdh_t),
\end{equation}
For the exponential family $\famF$, we consider the 
special family of multivariate Gaussian distributions.
More specifically, we let $q_t^{\backslash O}(\bdh_{t}) \triangleq 
q_t^{F}(\bdh_{t}) q_t^{R}(\bdh_{t}) \sim \CN(\bdm_t^{\backslash O}, 
\bdV_t^{\backslash O})$. The discrete distributions $q_t^O(\bds_t)$ are assumed 
to be the probability mass functions (pmf) of their corresponding random 
variables. Denoting  $\symsetK \triangleq 
\{ \bda_1, \bda_2, \cdots, \bda_{\M^K} \}$, the pmf can be defined as
\begin{align}
q_t^O(\bds_t) = \left[ P(\bds_t = \bda_1), 
\cdots, 
P(\bds_t = \bda_{\M^K}) \right]
\end{align}
Next, the hybrid posterior distribution is defined by
combining the 
$t^{th}$ factor in the likelihood function $p(\bdsetY | \bdsetS,
\bdsetH)$, namely $p(\bdy_t | \bds_t, \bdh_t)$, with \eqref{cavity_0} to get 
the following intermediate approximate posterior: 
\begin{align}
\hat{q}_t(\bds_t, \bdh_t) = \frac{p(\bds_t) q_t^{\backslash O}(\bdh_t) 
                                    p(\bdy_t | \bds_t, \bdh_t)}{Z_t},
\label{eq_int_approx_post}
\end{align}
where $Z_t$ is a normalization factor given by
\begin{align}
Z_t &= \sum_{\bds_t \in \symsetK} \int_{\bdh_t} p(\bds_t) 
              q_t^{\backslash O}(\bdh_t) p(\bdy_t | \bds_t, \bdh_t)d\bdh_t
                                                                    &\nonumber \\
    &= \sum_{\bds_t \in \symsetK} p(\bds_t) \times &
		\nonumber \\
		& \quad \int_{\bdh_t} 
              \CN (\bdh_t | \bdm_t^{\backslash O}, \bdV_t^{\backslash O}) 
	             \CN(\bdy_t | \bdS_t \bdh_t, \bdR_w)d\bdh_t
                                                                    &\nonumber \\
    &= \sum_{\bds_t \in \symsetK} p(\bds_t) \, 
              \CN(\bdy_t | \bdS_t \bdm_t^{\backslash O},\bdSigma_t),&
\label{z}
\end{align}
In \eqref{z}, $\bdSigma_t \triangleq \bdS_t \bdV_t^{\backslash O} \bdS_t^H + \bdR_w$. 
Next 
we project $\hat{q}_t(\bds_t, \bdh_t)$ onto the closest 
distribution (in the sense of Kullback-Leibler divergence) in $\famF$ to find 
$q_t(\bds_t, \bdh_t) = q_t(\bds_t)q_t(\bdh_t)$:
\begin{flalign}\label{KL_first}
q_t(\bds_t, \bdh_t) &= \argmin_{q_t(\bds_t, \bdh_t) \in {\famF}} 
  \KL \left( \hat{q}_t(\bds_t, \bdh_t) \| q_t(\bds_t, \bdh_t) \right)
& \nonumber \\
&= \argmin_{q_t(\bds_t), q_t(\bdh_t) \in {\famF}} 
  \KL \left(\hat{q}_t(\bds_t, \bdh_t) \| q_t(\bds_t) q(\bdh_t) \right), &
\end{flalign}
where $\KL(\cdot \| \cdot)$ denotes the Kullback-Leibler divergence. It is 
shown in \cite{Ghavami3} that the above optimization problem can be divided 
into the following two separate optimizations:
\begin{align}
\label{eq_KLh}
q_t(\bdh_t) &= \argmin_{q_t(\bdh_t) \in {\famF}} 
  \KL \left( \hat{q}_t(\bdh_t) \| q_t(\bdh_t) \right),
\\
\label{eq_KLs}
q_t(\bds_t) &= \argmin_{q_t({\bds}_t) \in {\famF}}
  \KL \left(\hat{q}_t({\bds}_t) \| q_t({\bds}_t) \right),
\end{align}
where $\hat{q}_t(\bdh_t)$ and $\hat{q}_t({\bds}_t)$ are the marginal distributions 
of $\bdh_t$ and $\bds_t$, respectively, 
derived from their joint distribution $\hat{q}_t({\bds_t},\bdh_t)$. 

The solution to \eqref{eq_KLh} is obtained from the so-called moment matching 
property \cite{Ghavami3}. Since the approximated posterior $q_t(\bdh_t) \in 
\famF$, we assume $q_t(\bdh_t) \sim \CN(\bdh_t | \bdm_t, \bdV_t)$. The moment matching property
implies
\begin{align}
\bdm_t &= \Exp_{\hat{q}_t(\bdh_t)}[\bdh_t],\label{mt_1}
\\
\bdV_t &= \Exp_{\hat{q}_t(\bdh_t)}[ \bdh_t \bdh_t^H] - 
                          \bdm_t \bdm_t^H,\label{Vt_1}
\end{align}
The values of $\bdm_t$ and $\bdV_t$ are given by the following 
lemma whose proof is omitted here.

\begin{lemma} \label{lemma1}
\begin{enumerate}
\item
The posterior mean value $\bdm_t$ is given by 
\begin{align} \label{eq_post_h_mean}
\bdm_t &= \bdm_t^{\backslash O}+ \bdV_t^{\backslash O} \nabla^H_m,
\end{align}
where 
\begin{align}\label{eq_Gradm}
\nabla^H_m &\triangleq \left(\frac{\partial}{\partial \bdm_t^{\backslash O}} \log Z_t\right)^H 
                                                                    \nonumber \\
         &= \frac{1}{Z_t} \sum_{\bds_t \in \symsetK} p(\bds_t) 
            \CN(\bdy_t | \bdS_t \bdm_t^{\backslash O}, \bdSigma_t) 
            \bdS_t^H \bdSigma_t^{-1} \bdzeta_t,
\end{align}
where $\bdzeta_t \triangleq \bdy_t - \bdS_t \bdm_t^{\backslash O}$ and $Z_t$ is 
given in \eqref{z}.
\item
The covariance matrix $\bdV_t$ is given by
\begin{align} \label{eq_post_h_cov}
\bdV_t = \bdV_t^{\backslash O} - \bdV_t^{\backslash O} 
          \left( \nabla_m^H \nabla_m -\nabla_V\right) \bdV_t^{\backslash O}
\end{align}
where  
\begin{align}
\nabla_V &\triangleq 
  \left(\frac{\partial \log Z_t}{\partial \bdV_t^{\backslash O}}\right)^T 
                                                                    \nonumber \\
  &= \frac{1}{Z_t}\sum_{\bds_t \in \symsetK} p(\bds_t)
     \CN \left( \bdy_t | \bdS_t \bdm_t^{\backslash O}, \bdSigma_t \right)\nonumber \times\\ 
  &\quad   \bdS_t^H \left( \bdSigma_t^{-1} \bdzeta_t \bdzeta_t^H \bdSigma_t^{-1}
      - \bdSigma_t^{-1} \right) \bdS_t
\label{eq_Gradv}
\end{align}
\end{enumerate}
Note that to simplify notation   
we have dropped the index $t$ from the left side of \eqref{eq_Gradm} and \eqref{eq_Gradv} here and in the following.
\end{lemma}


{Thus, $q_t^O(\bdh_t)$ can be updated as
\begin{equation}\label{update_q_O_ht}
q_t^O(\bdh_t)=\frac{q_t(\bdh_t)}{q^{\backslash O}_{t}(\bdh_t)}\propto \CN \left(\bdh_t|\bdm_t^O,\bdV^O_t\right).
\end{equation}
As discussed in \cite{Qi}, during the iterations of the algorithm,
$\bdV^O_t$ may become singular. Therefore, to avoid numerical issues 
we write the mean and covariance as the natural parameters as follows:
\begin{align}\label{mu_O_t}
\bdmu^O_t&=\bdLambda^O_t\bdm^O_t=\bdV^{-1}_t\bdm_t-\left(\bdV^{\backslash O}_t\right)^{-1}\bdm^{\backslash O}_t,\\
\bdLambda^O_t&=\left(\bdV^O_t\right)^{-1}=\bdV^{-1}_t-\left(\bdV^{\backslash O}_t\right)^{-1},\label{lambda_O_t}
\end{align}}
The solution to \eqref{eq_KLs} is to match the pmf of the posterior 
distribution $q_t(\bds_t)$ to the marginal $\hat{q}_t(\bds_t)$ which results in
\begin{equation} \label{eq_post_s_pmf}
  q_t(\bds_t)=
	\frac{p(\bds_t)}{Z_t} 
          \CN \left( \bdy_t | \bdS_t \bdm_t^{\backslash O}, \bdSigma_t \right),
\end{equation}
Therefore, $q^O_t(\bds_t)$ is updated by
\begin{equation}\label{st_est}
q^O_t(\bds_t)=\frac{q_t(\bds_t)}{p(\bds_t)}=\frac{1}{Z_t}
     \CN \left( \bdy_t | \bdS_t \bdm_t^{\backslash O}, \bdSigma_t \right),
\end{equation}

We now consider the beginning portion of the transmitted frame and for 
$t=1,\ldots,T_p$ for which we do not compute $q^O_t(\bds_t)$ as $\bds_t$ are the known pilot symbols. 
To update $q^O_t(\bdh_t)$, given $q^{\backslash O}_t(\bdh_t)$ the posterior factor $q_t(\bdh_t)$ 
can be directly updated from \eqref{eq_post_h_mean} and \eqref{eq_post_h_cov} 
by using $\nabla^H_m=\bdS_t^H \bdSigma_t^{-1} \bdzeta_t$ and 
$\nabla_V=\bdS_t^H \left( \bdSigma_t^{-1} \bdzeta_t \bdzeta_t^H \bdSigma_t^{-1} - \bdSigma_t^{-1} \right) \bdS_t$.
We should point out that for this part of the frame, the summations in
\eqref{z}, \eqref{eq_Gradm}, and \eqref{eq_Gradv} 
reduces to a single term corresponding to the known pilot symbol $\bds_t$. 
Following this, $q^O_t(\bdh_t)$ can be computed from \eqref{mu_O_t} and \eqref{lambda_O_t}.

\renewcommand{\baselinestretch}{1.2}

\begin{algorithm}\label{algo1}
  \footnotesize
\DontPrintSemicolon
\SetKwInput{KwPara}{Parameters}
\SetKwFor{ForEach}{for each}{}{end}
\SetKwRepeat{Repeat}{repeat}{until}
\SetKw{Break}{break}
\KwIn{$\bdsetY$}
\KwPara{$\sigma_w^2$, $\bdA$, $\bdQ$, $n$, $\epsilon$, $\bdm^{\backslash R}_0$, $\bdV^{\backslash R}_0$}
\KwOut{$\hat{\bdsetS}_d$, $\hat{\bdsetH}$}
\tcc{Initial filtering pass (KF-M run)}
\ForEach{$t = \{1,2,\ldots,T\}$}
{
Compute $\bdm^F_t$ and $\bdV^F_t$ via \eqref{F_factor} then set 
$\bdm^{\backslash O}_t=\bdm^F_t$ and $\bdV^{\backslash O}_t=\bdV^F_t$.\;
\uIf{$t>T_p$}
{Maximize $\CN \left(\bdy_t|\bdS_t\bdm^{\backslash O}_t,\bdSigma_t\right)$ for $\bds_t$ to get the maximizer $\hat{\bds}_t$.\\
Use $\hat{\bds}_t$ in \eqref{eq_Gradm_approx} and \eqref{eq_Gradv_approx} to compute $\bdm_t$ and $\bdV_t$ from \eqref{eq_post_h_mean} and \eqref{eq_post_h_cov}, 
respectively.}
\Else
{Use $\bds_t$ in \eqref{eq_Gradm_approx} and \eqref{eq_Gradv_approx} to compute $\bdm_t$ and $\bdV_t$ from \eqref{eq_post_h_mean} and \eqref{eq_post_h_cov}, 
respectively.}

Compute $\bdmu^{O}_t$ via \eqref{mu_O_t} and $\bdLambda^{O}_t$ via \eqref{lambda_O_t}.\\
Set $\bdm^{\backslash R}_t=\bdm_t$ and $\bdV^{\backslash R}_t=\bdV_t$.\\
}
\tcc{EP run}
\ForEach{$i = \{1,2,\ldots,n\}$}
{
\If{$i>1$}
{
\tcc{filtering pass}
\ForEach{$t = \{1,2,\ldots,T\}$}
{
Update $\bdm^F_t$ and $\bdV^F_t$ via \eqref{F_factor}.\;
For the followed up next smoothing pass, compute $\bdm^{\backslash R}_t$ via 
\eqref{mu_rem_R} and $\bdV^{\backslash R}_t$ via \eqref{cov_rem_R}.\;
}
}
\tcc{smoothing pass}
\ForEach{$t = \{T,T-1,\ldots,1\}$}
{
Set $\bdm_T=\bdm^{\backslash R}_T$ and $\bdV_T=\bdV^{\backslash R}_T$.\\
\If{$t<T$}
{Update $\bdm_t$ using \eqref{eq_post_h_mean_R} and $\bdV_t$ using \eqref{eq_post_h_cov_R}.\\
}
Compute $q^{\backslash O}_t(\bdh_t)=\frac{q_t(\bdh_t)}{q^O_t(\bdh_t)}\sim \CN 
\left(\bdh_t | \bdm^{\backslash O}_t,\bdV^{\backslash O}_t\right)$ as follows:\\
\begin{align}
\bdm^{\backslash O}_t&=\bdV^{\backslash O}_t\left(\bdV^{-1}_t\bdm_t-\bdmu^O_t\right),\nonumber \\
\bdV^{\backslash O}_t&=\left(\bdV^{-1}_t-\bdLambda^O_t\right)^{-1},\nonumber
\end{align}
\uIf {$t> T_p$}
{Maximize $\CN \left(\bdy_t|\bdS_t\bdm^{\backslash O}_t,\bdSigma_t\right)$ for $\bds_t$ to get the maximizer $\hat{\bds}_t$.\\

Use $\hat{\bds}_t$ in \eqref{eq_Gradm_approx} and \eqref{eq_Gradv_approx} to update 
$\bdm_t$ via \eqref{eq_post_h_mean} and $\bdV_t$ via \eqref{eq_post_h_cov}.}
\Else
{Use $\bds_t$ in \eqref{eq_Gradm_approx} and \eqref{eq_Gradv_approx} to update 
$\bdm_t$ via \eqref{eq_post_h_mean} and $\bdV_t$ via \eqref{eq_post_h_cov}.
}
Compute $\bdmu^{O}_t$ via \eqref{mu_O_t} and $\bdLambda^{O}_t$ via \eqref{lambda_O_t}.\\
}
\tcc{Check for convergence: Keep track of $\bdm_t$ for each $i^{th}$ iteration}
\If{$\frac{\Big|\Big|\bdm^{i}_t-\bdm^{i-1}_t\Big|\Big|}{\Big|\Big|\bdm^{i-1}_t\Big|\Big|}<\epsilon$}
{break;
}
}

Populate $\hat{\bdsetS}_d = \{ \hat{\bds}_{T_p+1}, \hat{\bds}_{T_p+2}, \ldots, \hat{\bds}_T \}$. \;
Populate $\hat{\bdsetH} = \{ \bdm_1, \bdm_2, \ldots, \bdm_T \}$. \;
\caption{Semi-blind EP Algorithm}
\end{algorithm}


\renewcommand{\baselinestretch}{1.7}


Next, we need to update $q_t^F(\bdh_t)$ and $q_t^R(\bdh_t)$ for the entire transmitted frame. Following the 
steps summarized in Fig. \ref{fig:Forward_Reverse_computation} we define the following intermediate distribution 
\begin{equation}\label{cavity_FR}
  r_t(\bdh_{t-1}, \bdh_t) \triangleq q_{t-1}^{\backslash R} (\bdh_{t-1}) 
                            p(\bdh_t | \bdh_{t-1}) q_t^{\backslash F}(\bdh_{t}),
\end{equation}
where $q_{t}^{\backslash R} (\bdh_t) \triangleq q_t(\bdh_t) / q_t^{R} (\bdh_t)$ 
and $q_{t}^{\backslash F} (\bdh_t) \triangleq q_t(\bdh_t) / q_t^{F} (\bdh_t)$ 
are the cavity distributions, given by
$\CN(\bdm_t^{\backslash i}, \bdV_t^{\backslash i})$ where $i = F$, $R$. Via some algebraic manipulations, 
we can show that 
\begin{align}
  r_t(\bdh_{t-1}, \bdh_t)&\propto\nonumber\\
		&\CN\left(
    \begin{bmatrix}
      \bdh_{t-1}\\
      \bdh_t
    \end{bmatrix} \middle|
    \begin{bmatrix}
      \bdmu_{t-1}\\
      \bdmu_t
    \end{bmatrix},
    \begin{bmatrix}
      \bdLambda_{t-1,t-1} & \bdLambda_{t-1,t}\\
      \bdLambda_{t,t-1} & \bdLambda_{t,t}\\
    \end{bmatrix}^{-1}
      \right), \label{eq_CNtrans}
\end{align}
where
\begin{align}
  \bdLambda_{t-1,t-1} &\triangleq \left(\bdV_{t-1}^{\backslash R}\right)^{-1} + 
                                    \bdA^H \bdQ^{-1} \bdA \label{eq_Lprev},\\
  \bdLambda_{t,t-1} = \bdLambda_{t-1,t}^H &\triangleq -\bdQ^{-1} \bdA, 
                                                        \label{eq_Ltrans} \\
  \bdLambda_{t,t} &\triangleq \left(\bdV_t^{\backslash F}\right)^{-1} + 
                                \bdQ^{-1} \label{eq_Lnow},
\end{align}
and the means are related by
\begin{align}
  \bdLambda_{t-1,t-1} \bdmu_{t-1} + \bdLambda_{t-1,t} \bdmu_{t} &= 
    \left( \bdV_{t-1}^{\backslash R}\right)^{-1} \bdm_{t-1}^{\backslash R}, 
    \label{eq_muprev}\\
  \bdLambda_{t,t} \bdmu_{t} + \bdLambda_{t,t-1} \bdmu_{t-1} &= 
    \left( \bdV_{t}^{\backslash F}\right)^{-1} \bdm_{t}^{\backslash F}, 
    \label{eq_munow}
\end{align}
Next we project $r_t(\bdh_{t-1}, \bdh_t)$ onto $\famF$ by minimizing the following KL divergence to get
\begin{align}
  &\left( q_t(\bdh_t), q_t(\bdh_{t-1}) \right)=\nonumber \\ 
    &\quad \argmin_{ q_t(\bdh_t), q_{t-1}(\bdh_{t-1}) \in \famF} 
    \KL \left( r_t(\bdh_{t-1},\bdh_{t}) \middle\| q_t(\bdh_t) 
                              q_{t-1}(\bdh_{t-1}) \right), 
															\label{eq_KLtrans}
\end{align}
As in the case of the optimization in \eqref{KL_first} which resulted in 
\eqref{eq_KLh} and \eqref{eq_KLs}, the above optimization problem can 
also be decomposed into two separate problems. 
In each one, we minimize the KL divergence between $q_k(\bdh_k)$ $\forall$ $k \in \{t,t-1\}$ and the 
respective marginal distribution 
obtained from $r_t(\bdh_{t-1}, \bdh_t)$ by integrating out the other variable. 
Setting $q_t(\bdh_t)$ to 
the marginal distribution in the 
KL optimization problem, $q^F_t(\bdh_t)$ can be updated from
\begin{align}\label{qF_factor}
q^F_t(\bdh_t)&=\frac{q_t(\bdh_t)}{q^{\backslash F}_t(\bdh_t)}=\int_{\bdh_{t-1}}{q^{\backslash R}_{t-1}(\bdh_{t-1})
			p(\bdh_t|\bdh_{t-1}) d\bdh_{t-1}}\\ \nonumber
			&\propto \CN \left(\bdh_t|\bdm^F_t,\bdV^F_t\right),
\end{align}
where 
\begin{equation}\label{F_factor}
\bdm^F_t=\bdA \bdm^{\backslash R}_{t-1}, \quad \bdV^F_t=\bdA \bdV^{\backslash R}_{t-1} \bdA^H +\bdQ,
\end{equation}
To update $q^R_t(\bdh_t)$, we follow the Kalman smoothing derivation to directly incorporate it into
$q_t(\bdh_t)$. Towards this, we first compute the term $q^{\backslash R}_{t}(\bdh_t)$ as follows
\begin{align}\label{qR_factor}
q^{\backslash R}_{t}(\bdh_t)&=q^F_t(\bdh_t)q^O_t(\bdh_t),\\ \nonumber
&=\CN \left(\bdh_t|\bdm^F_t,\bdV^F_t\right) \CN\left(\bdh_t|\bdm^O_t,\bdV^O_t\right), \\ \nonumber
&\propto \CN \left(\bdh_t|\bdm^{\backslash R}_t,\bdV^{\backslash R}_t\right),
\end{align}
where
\begin{align}\label{mu_rem_R}
\bdm^{\backslash R}_t&=\bdV^{\backslash R}_t\left(\left(\bdV^F_t\right)^{-1}\bdm^F_t+\bdmu^O_t\right),\\
\bdV^{\backslash R}_t&=\left(\left(\bdV^F_t\right)^{-1}+\bdLambda^O_t\right)^{-1},\label{cov_rem_R}
\end{align}

Next, since \eqref{eq_muprev} is related to the marginal distribution of $\bdh_{t-1}$, 
we obtain the mean of the posterior distribution $q_{t-1}(\bdh_{t-1})$ by solving 
for $\bdmu_{t-1}$. By substituting $\bdmu_t=\bdm_t$, $\bdLambda_{t-1,t-1}$ from 
\eqref{eq_Lprev}, and $\bdLambda_{t-1,t}$ from \eqref{eq_Ltrans}, it can be shown that 
\begin{align}
  \bdmu_{t-1} &= \bdLambda_{t-1,t-1}^{-1} \Big[ 
    \left( \bdV_{t-1}^{\backslash R}\right)^{-1} \bdm_{t-1}^{\backslash R} - 
    \bdLambda_{t-1,t} \bdm_{t} \Big] \label{eq_mean_rev}\\
  \bdV_{t-1,t-1} &= \bdLambda_{t-1,t-1}^{-1} + \bdLambda_{t-1,t-1}^{-1}
     \bdLambda_{t-1,t} \bdV_t \bdLambda_{t,t-1} \bdLambda_{t-1,t-1}^{-1},
     \label{eq_cov_rev}
\end{align}
Using results from Kalman smoothing the above can be reduced as follows.
\begin{align}
  \bdLambda_{t-1,t-1}^{-1} &= \bdV_{t-1}^{\backslash R} - \bdJ_{t-1} \bdF_{t-1} 
    \bdJ_{t-1}^H, \\
  \bdLambda_{t-1,t-1}^{-1} \bdLambda_{t-1,t} &= -\bdJ_{t-1},
  \intertext{where}
  \bdF_{t-1} &= \bdQ + \bdA \bdV_{t-1}^{\backslash R} \bdA^H, \label{eq_F} \\
  \bdJ_{t-1} &= \bdV_{t-1}^{\backslash R} \bdA^H \bdF_{t-1}^{-1}, \label{eq_J}
\end{align}
The update equations are obtained by substituting \eqref{eq_F} and \eqref{eq_J} 
into \eqref{eq_mean_rev} and \eqref{eq_cov_rev} and adjusting the notation to 
update the $t$th factors:
\begin{align}
  \bdm_t &= \bdm_t^{\backslash R} + \bdJ_t \left( \bdm_{t+1} - \bdA 
  \bdm_t^{\backslash R} \right), \label{eq_post_h_mean_R}\\
  \bdV_t &= \bdV_t^{\backslash R} + \bdJ_t \left( \bdV_{t+1} - 
    \bdF_t \right) \bdJ_t^H, \label{eq_post_h_cov_R}
\end{align}

This completes all the posterior updates for the EP iteration. This algorithm is 
summarized in Algorithm \ref{algo1}.

\subsection{Reducing computational complexity} \label{RCC}
 For $t=T_p+1,\ldots,T$, computation of \eqref{eq_Gradm} and \eqref{eq_Gradv} require summation of ${\cal M}^K$ terms 
which is computationally challenging. Since pilot symbols are transmitted for
$t=1, \cdots, T_p$, even in the first pass of the algorithm,
$\bdm^{\backslash O}_t$ provides a reasonably good estimate for $\bdh_t$ for $t=T_{p}+1, \cdots, T$. 
Therefore, the terms in 
$\bdV^{\backslash O}_t$ and $\bdSigma_t$ become smaller.
As a result, the PDF  
$\CN \left(\bdy_t|\bdS_t\bdm^{\backslash O}_t,\bdSigma_t\right)$ becomes narrow and 
all the summands in \eqref{eq_Gradm} and \eqref{eq_Gradv} become negligible except for the single term in which
$\bdy_t$ is close to $\bdS_t\bdm^{\backslash O}_t$. To find the dominant term we can use either one of the two
following methods:\\
a) MMSE estimator:
\begin{align}
\textbf{x}_t=\left(\left(\bdH^{\backslash O}_t\right)^H\bdR^{-1}_w\bdH^{\backslash O}_t 
+(E_s)^{-1}\bdI_K\right)^{-1}\left(\bdH^{\backslash O}_t\right)^H\bdR^{-1}_w\bdy_t, 
\end{align}
where $\bdH^{\backslash O}_t=\text{vec}^{-1}(\bdm^{\backslash O}_t)$,
where $\text{vec}^{-1}(.)$ is the inverse of the $\text{vec}(.)$ operation,
and followed by the hard decision
\begin{equation}
\hat{\bds}_t=\argmin_{{\bds}_t \in \symsetK} \mid \mid{\bds}_t-\textbf{x}_t\mid \mid,
\end{equation}
b) ML estimator :
\begin{equation}
\hat{\bds}_t=\argmax_{{\bds}_t \in \symsetK} \CN \left(\bdy_t|{\bdS}_t\bdm^{\backslash O}_t,\bdSigma_t\right),
\end{equation} 

Once $\hat{\bds}_t$ is computed from above, 
then \eqref{eq_Gradm} and \eqref{eq_Gradv} can be approximated with 
\begin{align}\label{eq_Gradm_approx}
\nabla^H_m &\approx \hat{\bdS}_t^H \bdSigma_t^{-1} \hat{\bdzeta}_t,
\end{align}
\begin{align}
\nabla_V &\approx \hat{\bdS}_t^H \left( \bdSigma_t^{-1} \hat{\bdzeta}_t \hat{\bdzeta}_t^H \bdSigma_t^{-1}
      - \bdSigma_t^{-1} \right) \hat{\bdS}_t,\label{eq_Gradv_approx}
\end{align}
where $\hat{\bdS}_t = \hat{\bds}_t^T \otimes \bdI_M$ and $\hat{\bdzeta}_t=\bdy_t-\hat{\bdS}_t\bdm^{\backslash O}_t$. 
\begin{remark}
With the above simplification, the computational complexity of our 
algorithm is dominated by \eqref{eq_post_h_mean}, \eqref{eq_post_h_cov}, 
\eqref{mu_O_t}, \eqref{lambda_O_t}, \eqref{F_factor}, \eqref{mu_rem_R}, 
\eqref{cov_rem_R}, \eqref{eq_post_h_mean_R}, \eqref{eq_post_h_cov_R}, 
and computation of $\bdm^{\backslash O}_t$, $\bdV^{\backslash O}_t$ in 
the smoothing pass. 
Thus, the computational complexity of our EP algorithm is $O(nT(M^3K^3+M^2K^2))$.
This complexity is the same as that of 
the conventional Kalman fitering and smoothing algorithms.
\end{remark}
\section{Simulation Results}\label{s-results}

In this section, we evaluate the performance of the proposed {semi-blind} EP algorithm for joint channel estimation and symbol detection through simulations. 
We consider a cellular system with $L=4$ cells and $K=8$ users in each cell. 
The performance of the algorithms in the first cell is presented. 
The large-scale fading coefficients for the $K$ users in the cells are set to 
$\beta_{11k}=1$ and $\beta_{1ik}=a$, $k=1,\dots,K$ and $i=2,3,4$. 
The constant scalar $a$ models the cross gain between the first cell BS and the users in other cells \cite{Ngo_PC, Ngo_PC1}. 
By varying the value of $a$ 
we study the effect of pilot contamination and inter-cell interference
on the channel estimation and symbol detection.

{
Using the Kronecker model,
\cite{Kronecker_SC, Ghavami1} for the spatial correlation matrices $\bdR_{1ik}$, we assume
$[\bdR_{1ik}]_{m,n}=r(m-n)$, $m,n=1,2,\ldots,M$ and 
$i=1,2,3,4$, where $r(m)=(\rho)^{|m|}$}\footnote{For ease of presentation, 
we assume that all users have the same 
spatial correlation modeled by the parameter $\rho$. 
This assumption is valid when the angle of visibility to the 
target BS for the group of users in the neighboring cells 
is either the same or a mirror image of the angle of visibility for 
the group of user in the target cell \cite{massivemimobook, user_grouping}.}.
The transmitted frame of length $T$ is composed of $T_p$ 
pilots symbols in the beginning followed by the $T_d$ unknown data symbols. 
QPSK modulation with average 
symbol energy $E_s=0$ dB is assumed for both pilot symbols and the data symbols. 
Hadamard code is employed to ensure that the pilot symbols of users 
in the first cell are orthogonal.
Note that with our assumed signal model in
\eqref{eq_VEC}, {orthogonality between the pilot sequences in the 
first and neighboring cells is not assumed.}
The time-varying channel vectors for all the users in the first cell are 
generated according to \eqref{AR} 
with an initial Gaussian prior distribution on $\bdh_0$ 
with zero mean and covariance matrix $\bdR_h$. 
Further, it is assumed that
all the users in the first cell have the same normalized 
Doppler shift $f_d$ and therefore the diagonal components of matrix $\bdA$ in \eqref{AR} 
are set to $J_0(2\pi f_d)$\footnote{Note 
that increasing $f_d$ decreases $a_n$ and thus the temporal correlation among the channel vectors.}.

The proposed algorithm described in Algorithm \ref{algo1} is initialized with 
$\bdm^{\backslash R}_0=\textbf{0}$ and $\bdV^{\backslash R}_0=\bdR_h$.
Further, since EP 
converges in a few iterations, we set the maximum number of 
iterations to $n=10$ and the error tolerance between two iterations for terminating the algorithm to $\epsilon=10^{-6}$. 

The symbol error rate (SER) is averaged over all the users in the first cell and the 
channel estimation accuracy is measured using the following normalized error,
\begin{equation}
\text{$\delta_h$(dB)}=10\log_{10}\left[\frac{1}{T}\sum^T_{t=1}\frac{\Exp\left[\mid \mid \bdh_t-\hat{\bdh}_t\mid \mid ^2\right]}{\Exp\left[\mid \mid \bdh_t\mid \mid ^2\right]}\right],
\end{equation}

We note that \eqref{eq_Gradm_approx} and \eqref{eq_Gradv_approx}, together with
\eqref{F_factor}, \eqref{eq_post_h_mean}, 
and \eqref{eq_post_h_cov} 
represent the prediction and time-update equations of the Kalman filtering algorithm 
where the unknown data symbols are estimated from density maximization. 
Therefore we refer to the initial forward pass of our algorithm as the modified Kalman filter (KF-M). 
The performance of this semi-blind version of Kalman filter which emerges 
from our EP derivations is also presented. 
Further, \eqref{eq_post_h_mean_R} and 
\eqref{eq_post_h_cov_R} represent the backward recursion equations of Kalman smoother. 
Thus, the performance of the Kalman smoothing algorithm followed by a 
{single pass of KF-M}
is also presented here for comparison and is denoted by KS-M.
To benchmark the performance of semi-blind KF-M, KS-M, and EP, 
we also present the performance of the Kalman filter and smoother 
in a pure training mode (TM) 
when the entire frame is composed of known pilot symbols and only channel estimation is performed. 
These two cases, are referred to as KF-TM and KS-TM.
In channel estimation, KF-TM provides a lower bound for KF-M, 
and KS-TM provides a lower bound for KS-M, and EP.
Finally, we also plot the SER performance of
the MMSE estimator with known CSI (denoted PCSI) for comparison 
with SER performance of the proposed algorithms.
\begin{figure*}[ht]
    \centering
    \begin{minipage}{0.48\textwidth}
     	\includegraphics[width=0.99\textwidth]{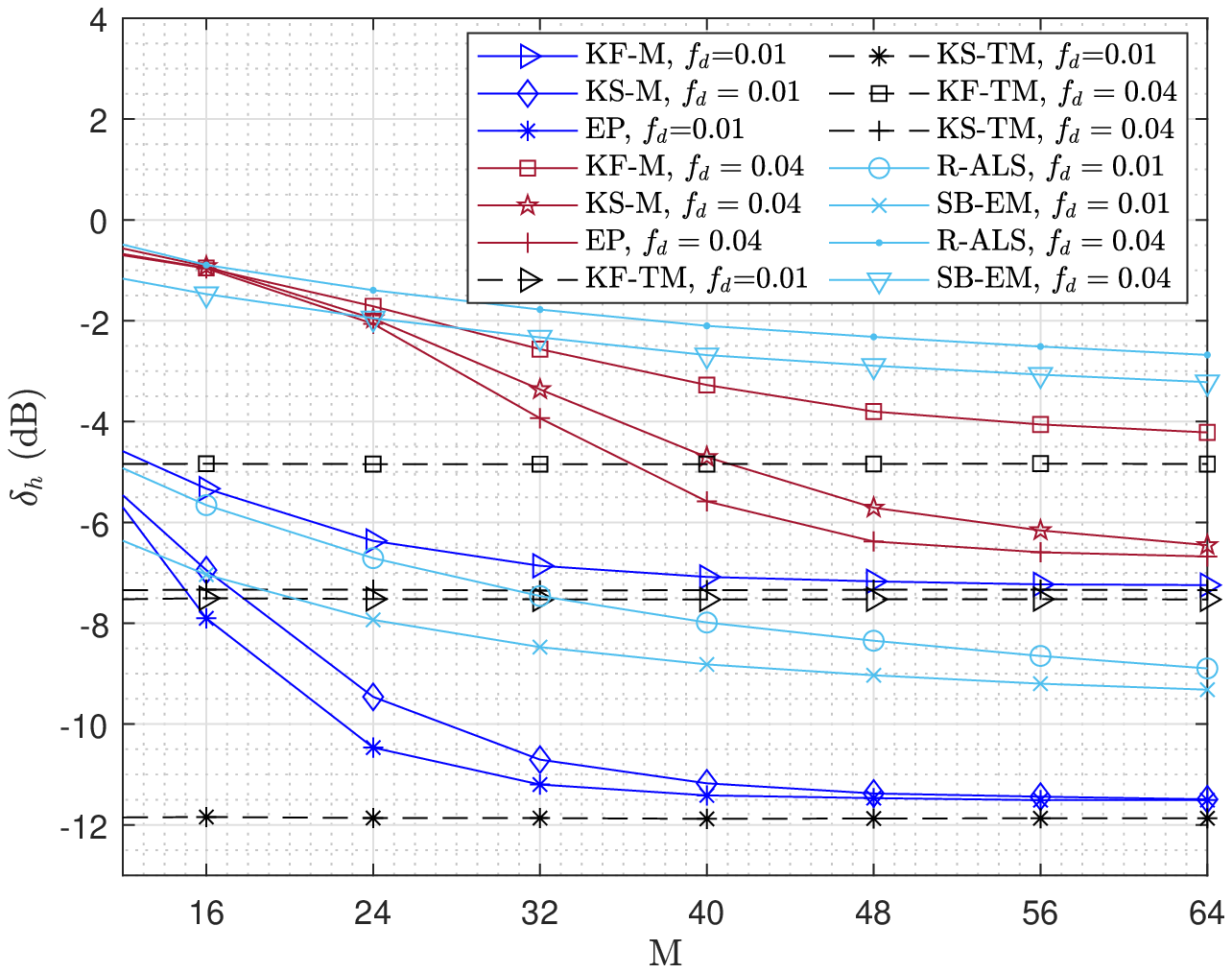}
	\caption{Channel estimation error versus the receiver's antenna array size $M$ and with parameters: $K=8$, $T_d=64$, $T_p=K$, $a=0.1$, $\rho=0$.}
	\label{fig:MSE_vs_M}
    \end{minipage}\hspace{.01\linewidth}
    \begin{minipage}{0.48\textwidth}
\includegraphics[width=0.99\textwidth]{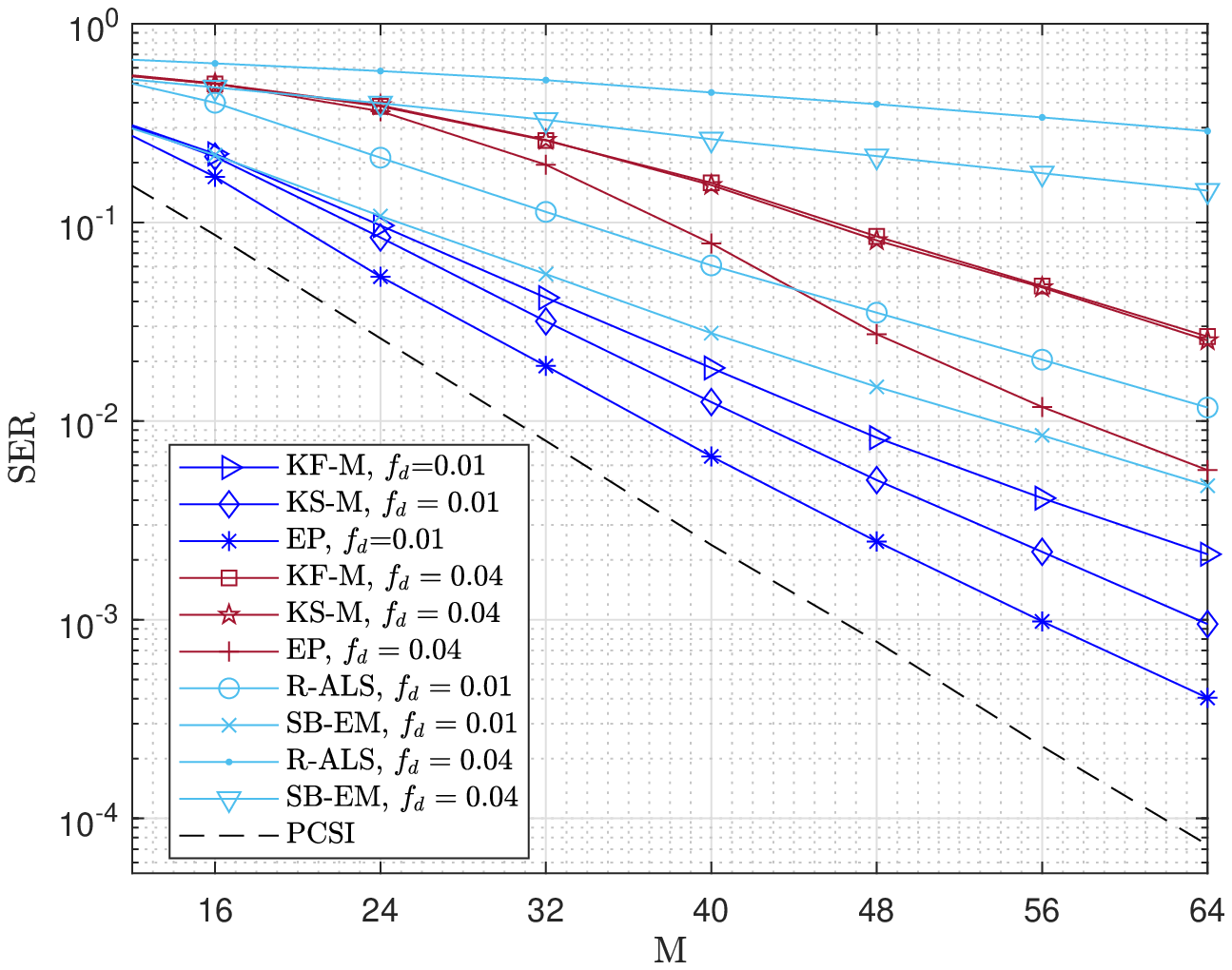}
	\caption{SER versus the receiver's antenna array size $M$ and with parameters: $K=8$, $T_d=64$, $T_p=K$, $a=0.1$, $\rho=0$.}
	\label{fig:SER_vs_M}
		\end{minipage}
\end{figure*}
\begin{figure*}[ht]
    \centering
    \begin{minipage}{0.48\textwidth}
     	\includegraphics[width=0.99\textwidth]{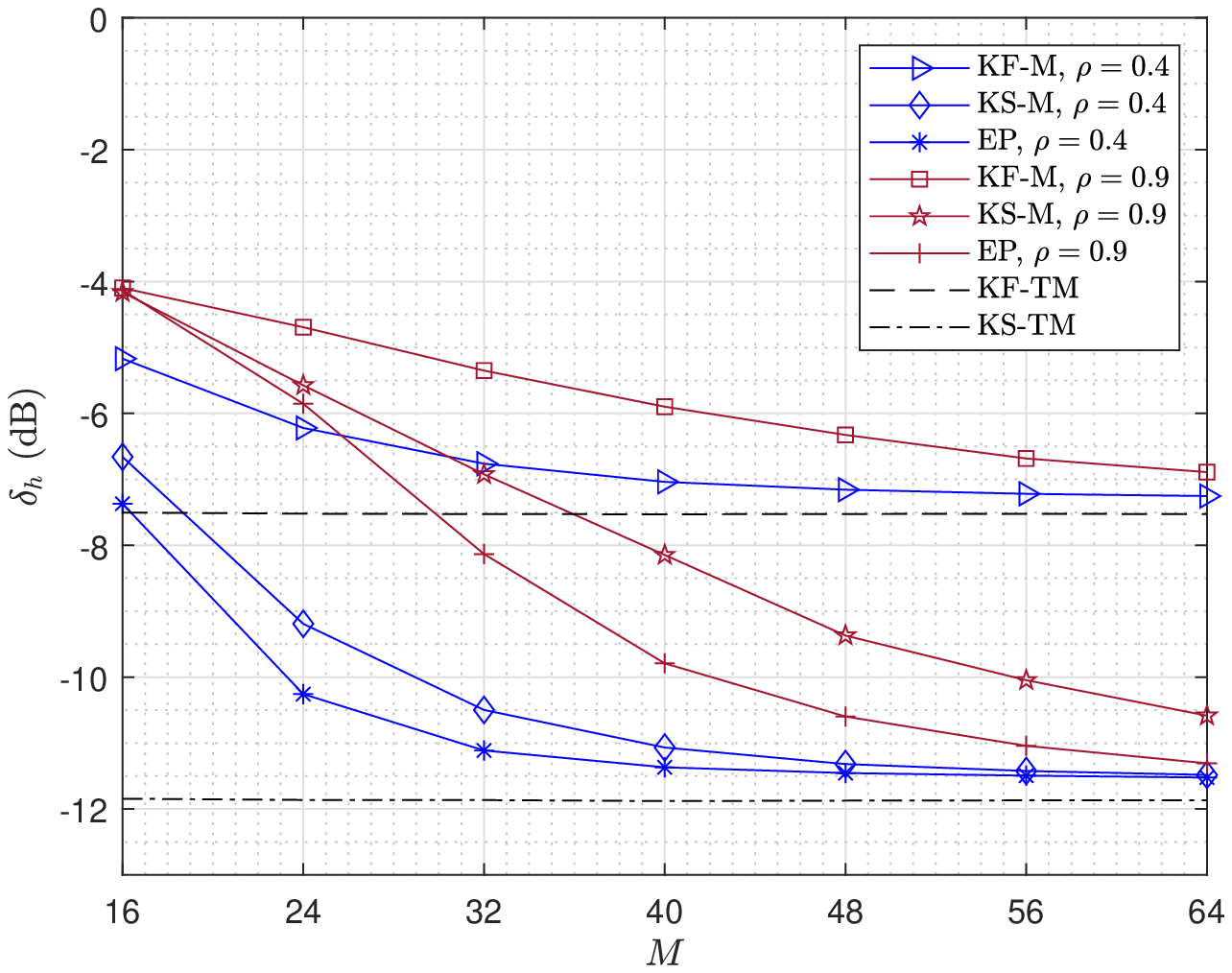}
	\caption{Channel estimation error versus the receiver's antenna array size $M$ and with parameters: $K=8$, $T_d=64$, $T_p=K$, $a=0.1$, $f_d=0.01$.}
	\label{fig:MSE_vs_M_vs_pho}
    \end{minipage}\hspace{.01\linewidth}
    \begin{minipage}{0.48\textwidth}
\includegraphics[width=0.99\textwidth]{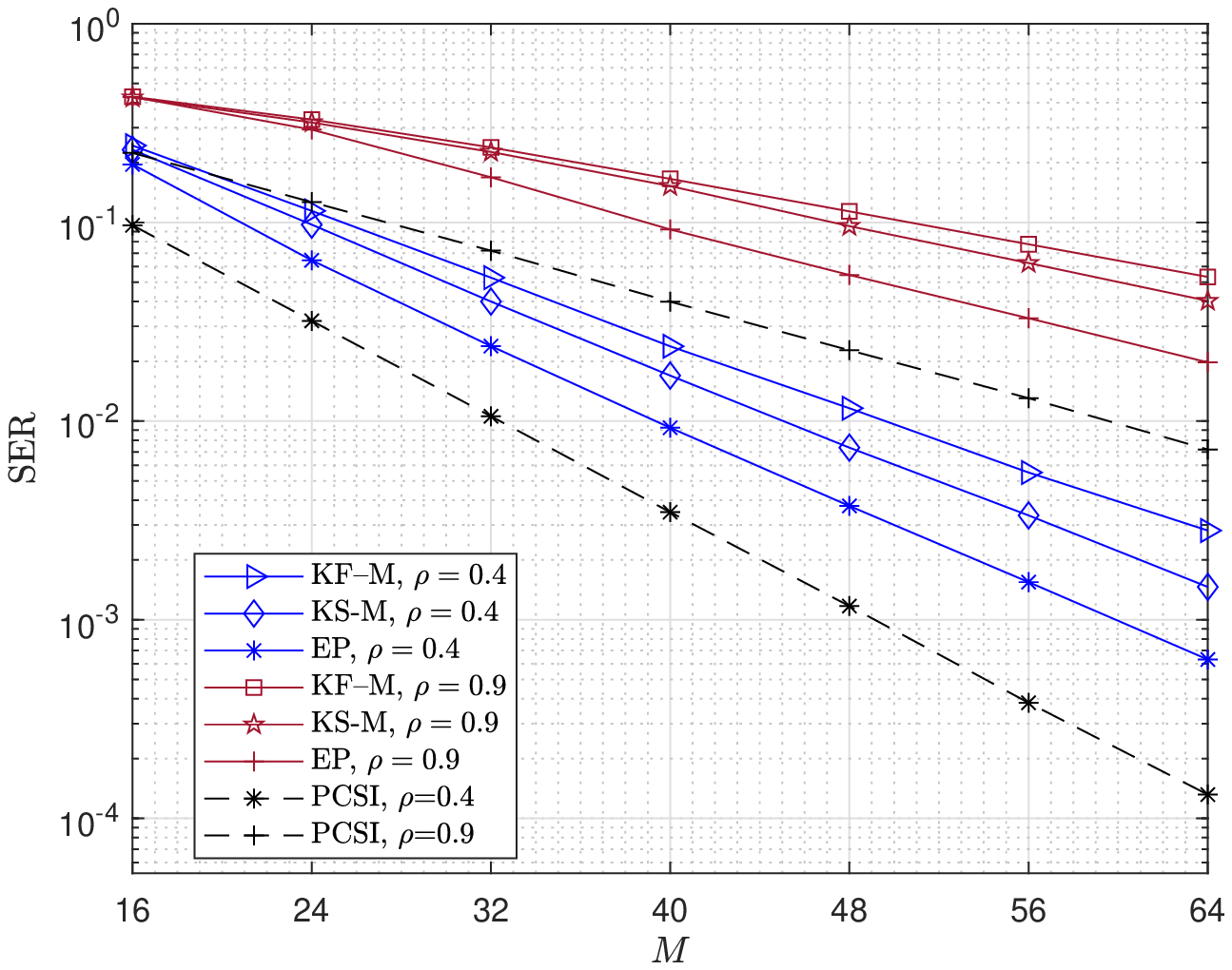}
	\caption{SER versus the receiver's antenna array size $M$ and with parameters: $K=8$, $T_d=64$, $T_p=K$, $a=0.1$, $f_d=0.01$.}
	\label{fig:SER_vs_M_vs_pho}
		\end{minipage}
\end{figure*}
\begin{figure*}
	\centering
	\begin{minipage}{0.48\textwidth}
				\includegraphics[width=0.99\textwidth]{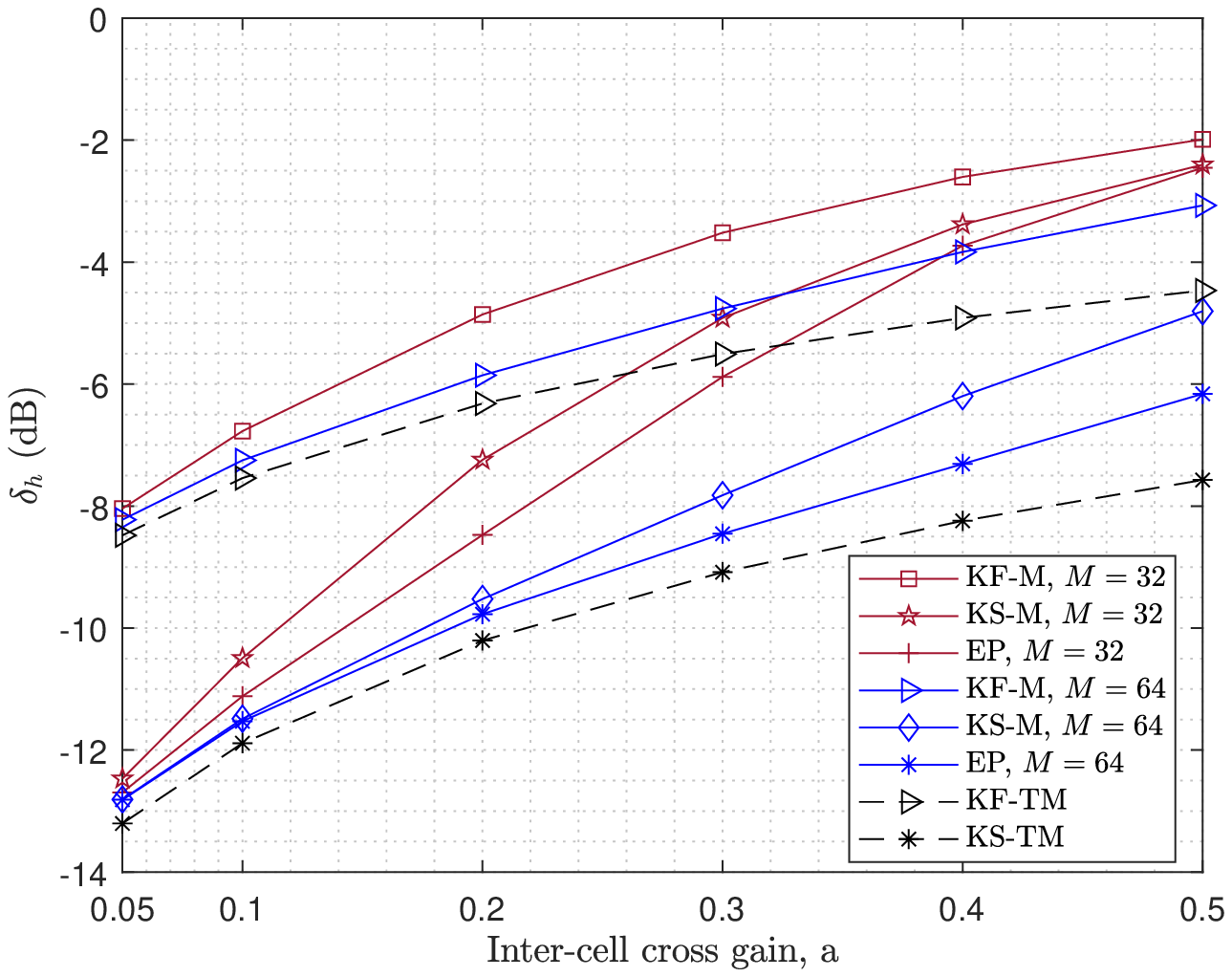}
	\caption{Channel estimation error versus the cross gain $a$ of users in other cells and with parameters: $K=8$, $T_d=64$, $T_p=K$, $f_d=0.01$, $\rho=0.4$.}
	\label{fig:MSE_vs_a}
   \end{minipage}\hspace{.01\linewidth}
    \begin{minipage}{0.48\textwidth}
	\includegraphics[width=0.99\textwidth]{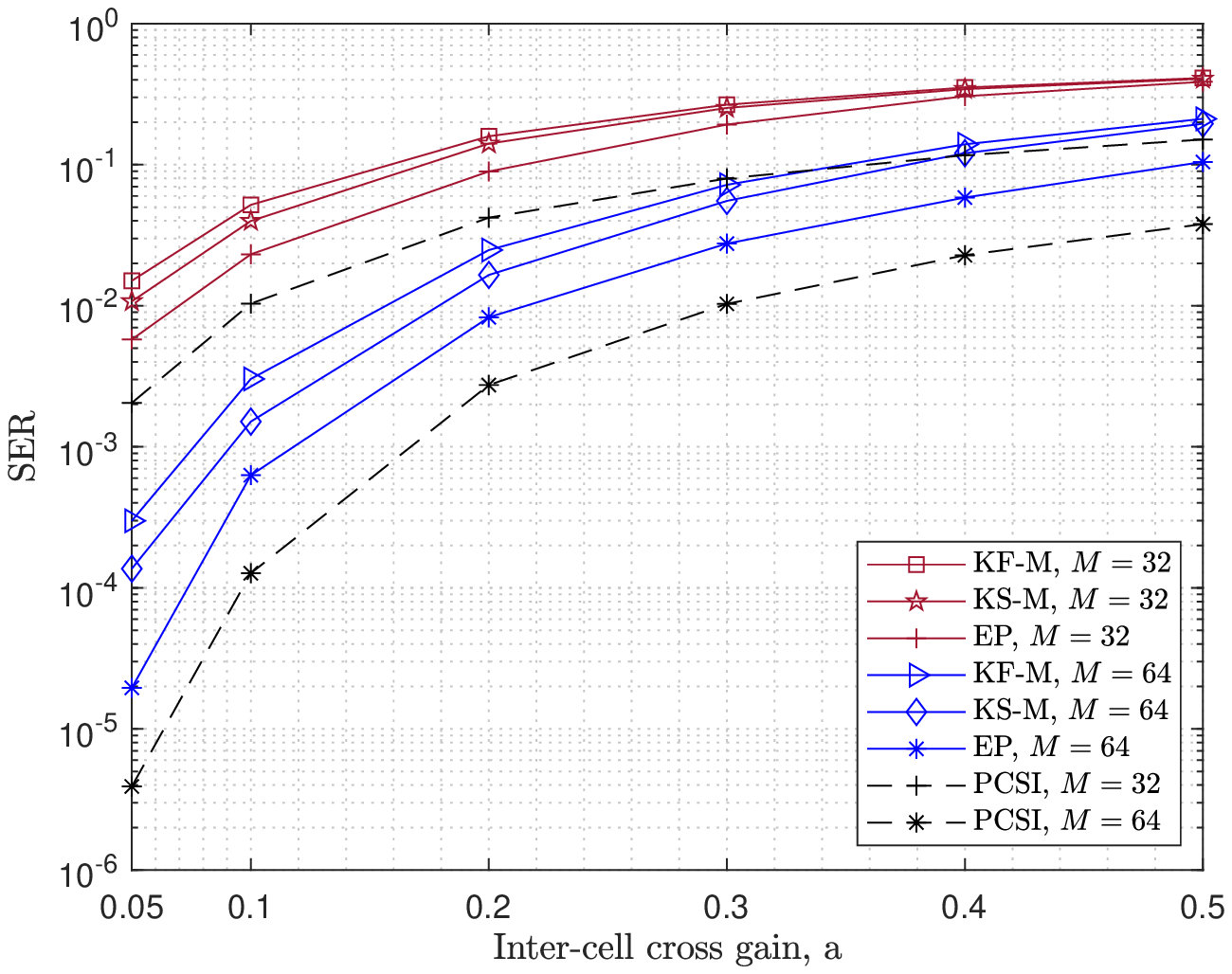}
	\caption{SER versus the cross gain $a$ of users in other cells and with parameters: $K=8$, $T_d=64$, $T_p=K$, $f_d=0.01$, 
	$\rho=0.4$.}
	\label{fig:SER_vs_a}
\end{minipage}
\end{figure*}

Fig. \ref{fig:MSE_vs_M} shows the channel estimation error versus 
the number of antennas $M$ for KF-M, KS-M, and EP algorithms in a 
spatially uncorrelated ($\rho=0$) but temporally correlated channel. 
We consider two different cases for the temporal 
correlation corresponding to $f_d=0.01$ and $f_d=0.04$\footnote{{
We should point out that these are the normalized values for the Doppler shift.
In other words, $f_d=f_D T_s$ where $f_D$ is the Doppler shift in  
Hz and $T_s$ is the symbol period in sec. For example for 
an OFDM system with a bit rate of $15$ Kbps per subcarrier,
resulting in a symbol rate of $7.5$ Ksps using QPSK modulation, we get $f_D=75$ Hz 
and $f_D= 300$ Hz for values of $f_d=0.01$ and $0.04$, respectively.
For a carrier frequency of $2$ GHz this implies a mobile velocity of
$40.5$ and $162$ Km/h, respectively. \label{foot.1}}}.
We observe that in both of these cases the performance of the algorithms improves 
with $M$ and EP has a significant improvement over KF-M. 
The channel is more time-varying in the case of $f_d=0.04$ and 
the estimation error is higher in this case.
The channel estimation error of KF-M approaches that of KF-TM for larger values of $M$
and the performance of EP approaches that of KS-TM. 
For both values of $f_d$, with increasing $M$ our proposed semi-blind EP algorithm converges faster to KS-TM than 
the semi-blind KS-M 
resulting in more accurate channel estimation. 
Note that while KF-TM and KS-TM algorithms employ $T=72$ pilot symbols to 
estimate the channel,
KF-M, KS-M, and EP use only $T_p=K=8$ pilot symbols to estimate 
the channel (as well as 
detect $T_d=64$ data symbols).
Channel estimation improvement with $M$
is a result of the so-called \textit{favorable propagation condition}
where the channel vectors of different users become mutually orthogonal as $M \longrightarrow \infty$.  
As a result, the performance of MMSE symbol estimator used in KF-M and 
EP improves. This in turn improves the 
channel estimation accuracy for KF-M, KS-M, and EP.
{
Fig. \ref{fig:MSE_vs_M} also shows the performance 
of the semi-blind expectation maximization (SB-EM) and the
regularized 
alternating least-square (R-ALS) algorithms proposed in  
\cite{Nayebi} and \cite{Liang_Semiblind}, respectively, 
for the channel described above using $T_p=8$ pilot symbols. 
We observe that for $f_d= 0.01$, 
both SB-EM and R-ALS perform better than the KF-M and KF-TM algorithms, but 
worse than the KS-M and EP. The 
improvement in performance over KF-M and KF-TM 
is because both SB-EM and R-ALS estimate the channel using 
the entire received frame, whereas at any given time, 
KF-M and KF-TM update the 
channel estimates using the received signals up to the present time.
Since KS-M, KS-TM, and EP use the entire received frame as well in a smoothing pass, 
they outperform SB-EM and R-ALS due to the underlying block-fading assumption 
of the latter two algorithms. 
Further, we observe that 
in the case of $f_d=0.04$, where the channel is highly time-varying,
the performance of R-ALS and SB-EM is worse than 
all the other algorithms.}

Fig. \ref{fig:SER_vs_M} depicts the SER performance of KF-M, KS-M, EP
and PCSI versus
the number of antennas $M$.   
We can see that for both values of $f_d$,
the SER performance of all the algorithms improves with 
$M$ and EP outperforms all the other algorithms except for PCSI (MMSE with known channel 
coefficients). 
Moreover, the improvement of EP over the other algorithms increases with $M$.
{The SER performance of SB-EM and R-ALS is also shown in Fig. \ref{fig:SER_vs_M}.
It can be seen from Figs. \ref{fig:MSE_vs_M} and \ref{fig:SER_vs_M} that 
the performance of algorithms developed under
the block-fading assumption is significantly degraded when 
the algorithms are applied to time-varying channels
as compared with algorithms specifically designed for such channels.}
 
\begin{figure*}[ht]
    \centering
    \begin{minipage}{0.48\textwidth}
     	\includegraphics[width=0.99\textwidth]{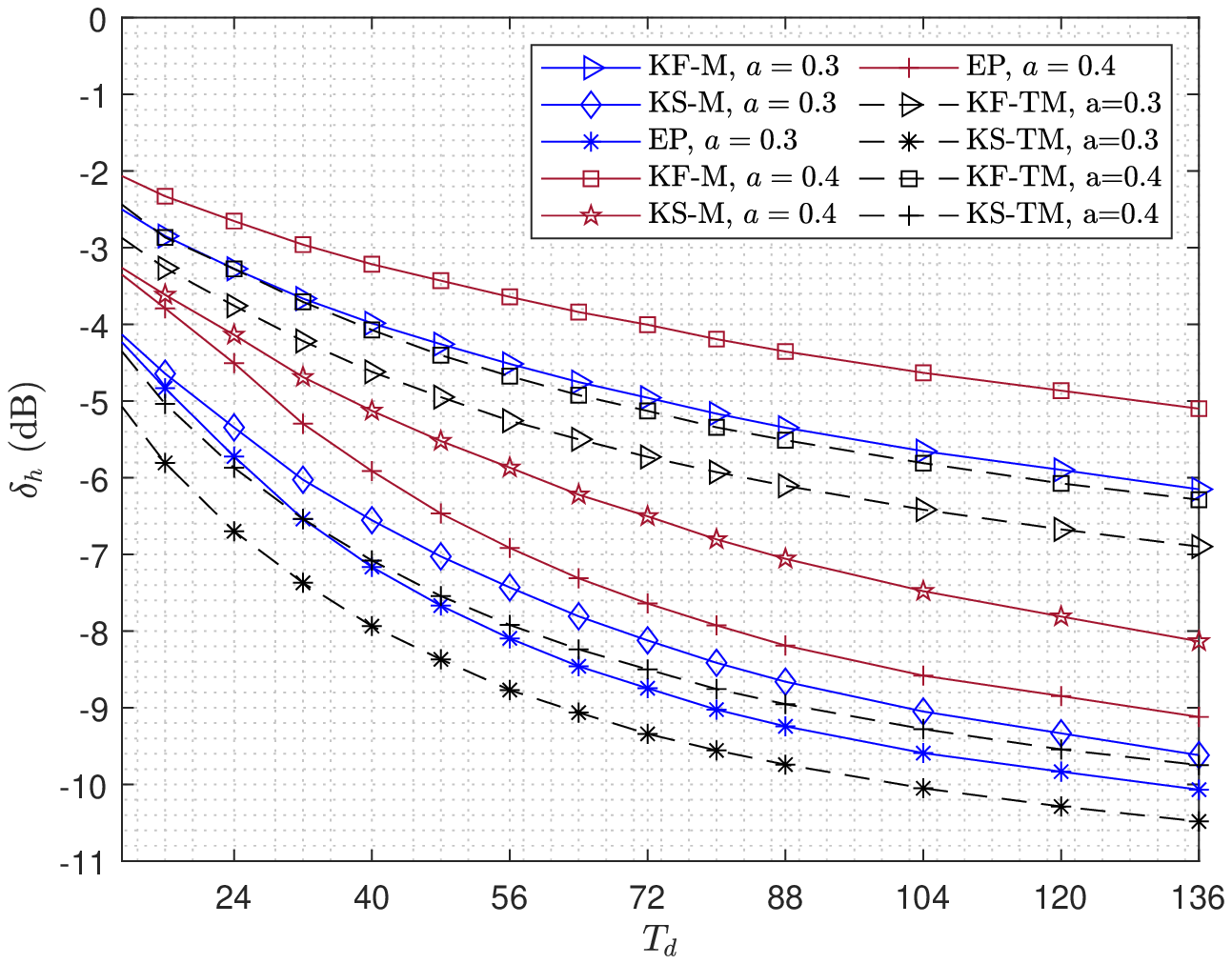}
	\caption{Channel estimation error versus $T_d$ and with parameters: $K=8$, $M=64$, $T_p=K$, $f_d=0.01$, $\rho=0.4$.}
	\label{fig:MSE_vs_Td}
    \end{minipage}\hspace{.01\linewidth}
    \begin{minipage}{0.48\textwidth}
\includegraphics[width=0.99\textwidth]{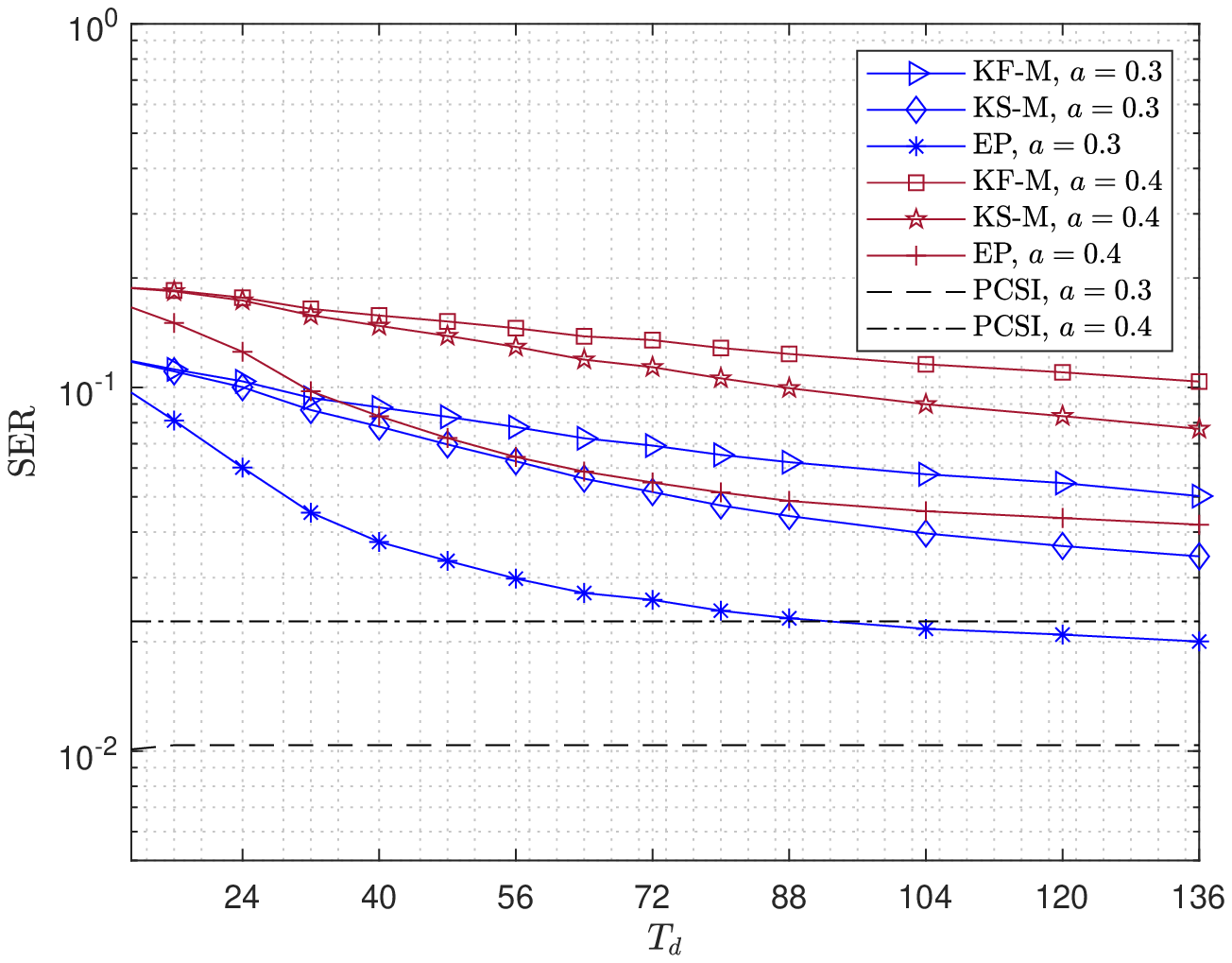}
	\caption{SER versus $T_d$ and with parameters: $K=8$, $M=64$, $T_p=K$, $f_d=0.01$, 
	$\rho=0.4$.}
	\label{fig:SER_vs_Td}
		\end{minipage}
\end{figure*}
\begin{figure*}
	\centering
	\begin{minipage}{0.48\textwidth}
				\includegraphics[width=0.99\textwidth]{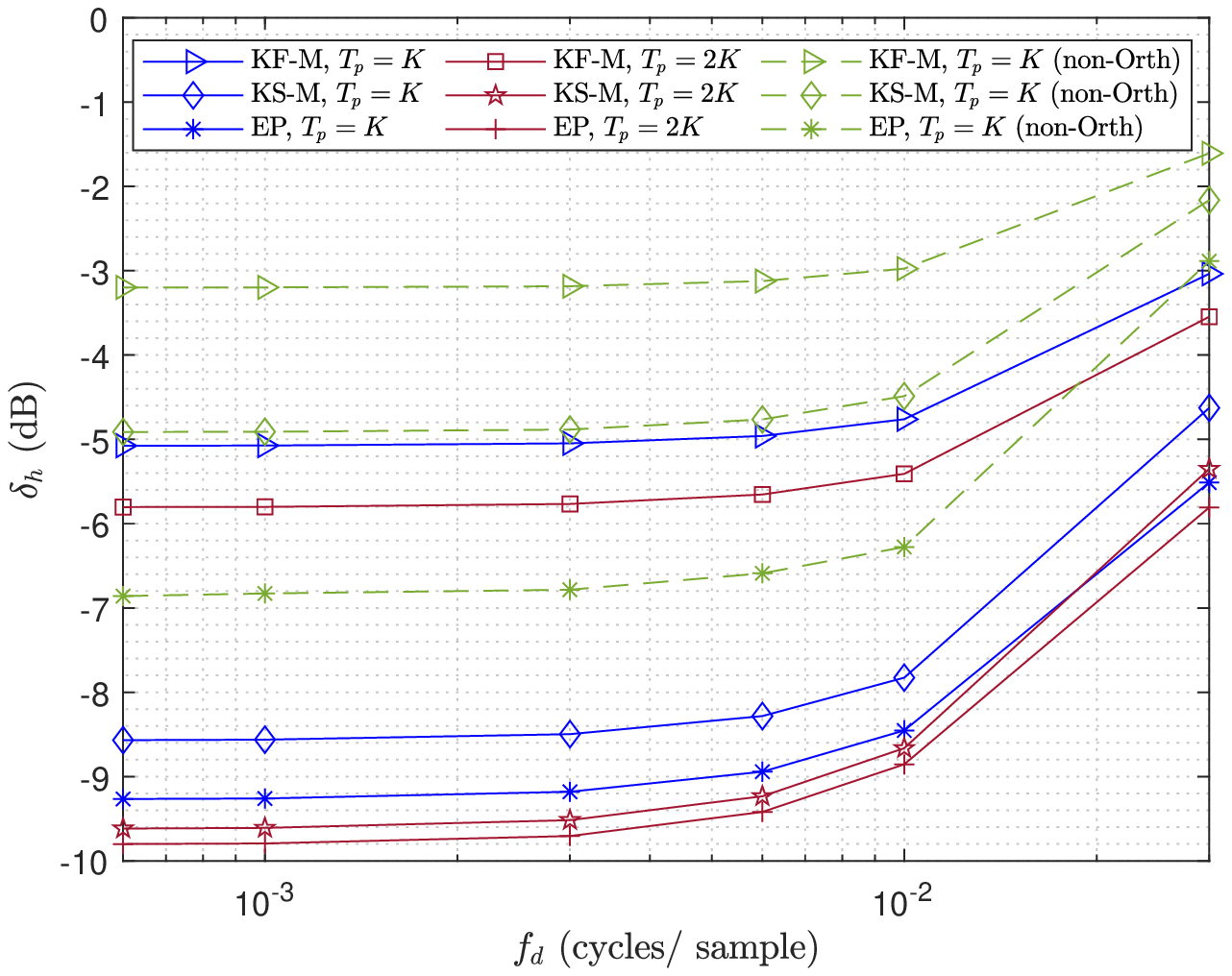}
	\caption{Channel estimation error versus the normalized Doppler shift $f_d$ and with parameters: $K=8$, $M=64$, $T_d=64$, 
	$\rho=0.4$, $a=0.3$.}
	\label{fig:MSE_vs_fd}
   \end{minipage}\hspace{.01\linewidth}
    \begin{minipage}{0.48\textwidth}
	\includegraphics[width=0.99\textwidth]{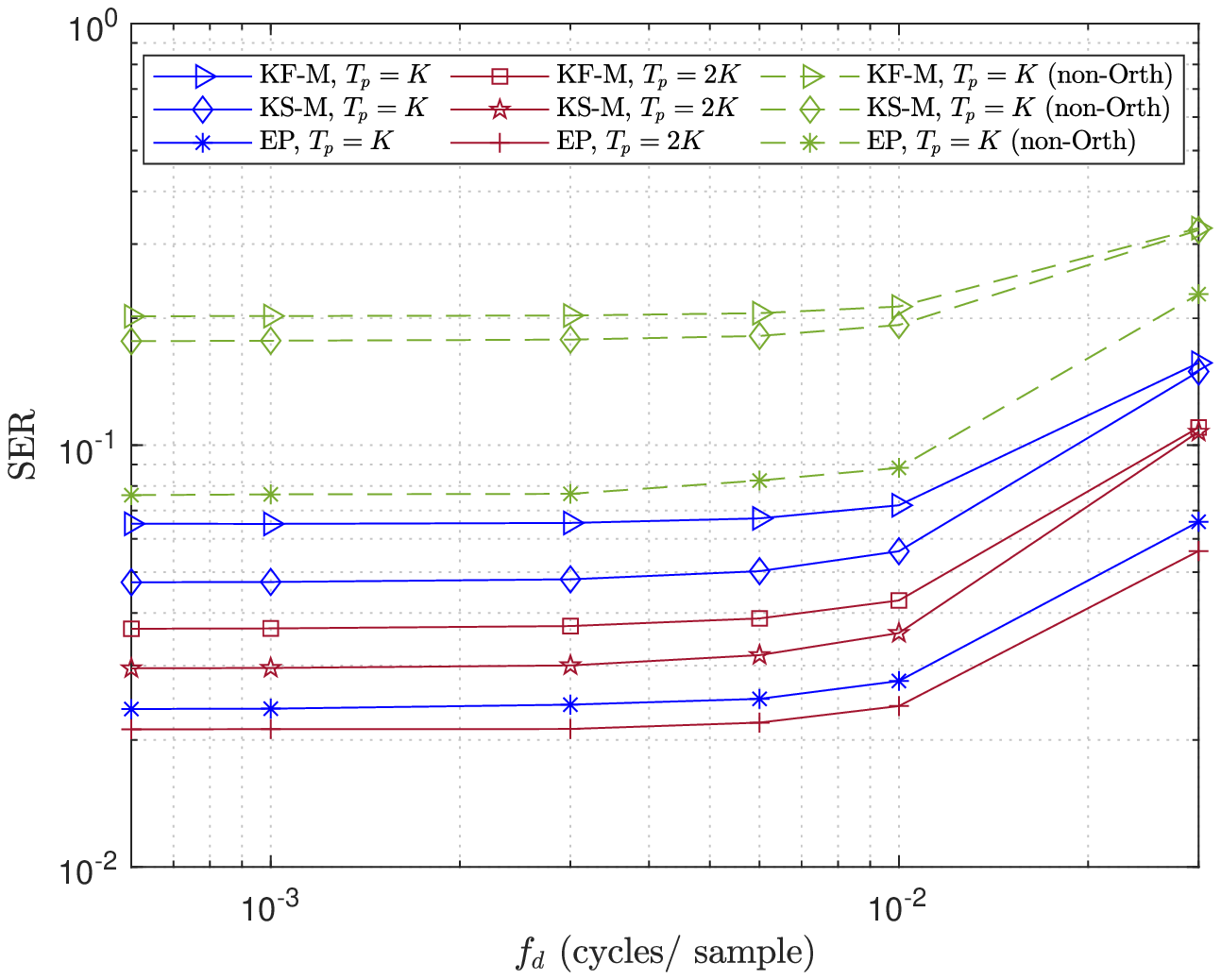}
	\caption{SER versus the normalized Doppler shift $f_d$ and with parameters: $K=8$, $M=64$, $T_d=64$, $\rho=0.4$, $a=0.3$.}
	\label{fig:SER_vs_fd}
\end{minipage}
\end{figure*}


Figs. \ref{fig:MSE_vs_M_vs_pho} and \ref{fig:SER_vs_M_vs_pho} 
show the
performance of KF-M, KS-M, and EP versus the number of antennas $M$ for a spatially and temporally correlated 
massive MIMO channel. For the case of $f_d=0.01$ we consider two different cases 
of spatial correlation corresponding to $\rho=0.4$ and $\rho=0.9$. 
We can see that 
in both cases the performance of the algorithms again improves with $M$ 
and EP significantly outperforms KF-M and KS-M. 
{A comparison of the three cases for $\rho=0$ (in Figs. \ref{fig:MSE_vs_M}
and \ref{fig:SER_vs_M}), $\rho=0.4$ and $\rho=0.9$ confirms the expected result
that as $\rho$ increases, channel diversity decreases resulting in 
the degradation of system performance. 
In addition, spatial channel correlation reduced
the level of channel hardening.
In conclusion, as spatial correlation increases,
a larger number of antennas are required to achieve the same level of performance
\cite{Bjornson}.}

In Figs. \ref{fig:MSE_vs_a} and \ref{fig:SER_vs_a}, 
we study the effect of pilot contamination on 
channel estimation error and SER. To this end, we 
vary the inter-cell cross gain $a$ between the BS in the first cell 
and the users in the three neighboring cells for the cases of
$M=32, 64$. As expected, as the 
cross gain $a$ increases, the performance of the algorithms degrades. 
However, as the figures show, EP significantly outperforms KF-M and KS-M.
Moreover, there is a significant improvement as $M$ increases from $32$ to $64$.

Figs. \ref{fig:MSE_vs_Td} and \ref{fig:SER_vs_Td} show the performance of 
the algorithms
versus the number of data symbols $T_d$ in the transmitted frame. 
Two different cases of inter-cell interference described by $a=0.3$ and $a=0.4$
are considered. It is observed 
that as $T_d$ increases, the channel estimation error and SER performance of all algorithms improves. 
This improvement is due to the fact that the semi-blind approach 
uses the $T_d$ data symbols as virtual pilot symbols in 
estimating the channel. Thus for a fixed 
number of antennas $M$, using a large $T_d$ 
can mitigate the impact of pilot contamination \cite{Nayebi}.
In these figures, the improvement in EP's performance is better than both KF-M and KS-M. 
This is due to the fact that EP 
updates the channel estimate at each 
time instant by incorporating the forward, reverse, and observation messages. 
In contrast, KF-M 
does not include the reverse and observation messages and KS-M 
does not include the observation messages.
Hence as $T_d$
increases, the channel 
estimates found through the smoothing pass of EP are more accurate 
than those from KS-M 
and KF-M. Further, iterative forward and backward pass
on the entire transmitted frame improves the SER performance of EP as well.

Finally, in Figs. \ref{fig:MSE_vs_fd} and \ref{fig:SER_vs_fd}, we study the 
performance of KF-M, KS-M, and EP versus the normalized Doppler shift $f_d$ for the
pilot sequence of length $T_p=K$ and $T_p=2K$.
It is seen that as $f_d$ increases (the temporal correlation among the channel vectors decreases), the 
performance of all algorithms degrades.
{It is interesting to note that for $f_d \leq 0.005$, 
the performance of the algorithms is almost constant with $f_d$.
In essence for these values of $f_d$, the channel may be assumed to be time-invariant (block fading). For the parameters listed in \footref{foot.1}, this translates to
mobile velocities less than $20$ Km/h.}
In these figures we also 
show the performance of the algorithms when non-orthogonal pilot sequences 
for users in the first cell are used.
In this case for each user $T_p$ QPSK symbols are randomly generated 
and used at the beginning of each frame as the pilot symbols.
Although EP still outperforms KF-M and KS-M in this case, a degradation in 
performance is observed compared to the previous case for all the algorithms.


\section{Conclusions}\label{conclusn}

In this paper, we propose a semi-blind expectation propagation (EP) based algorithm for joint channel 
estimation and symbol detection in the uplink of multi-cell massive MIMO systems for spatially and temporally 
correlated channels. 
EP algorithm is developed to approximate the a posteriori distribution of 
the channel matrix and data symbols with a distribution from an exponential family. 
The latter is then used to directly estimate the channel and detect the data symbols. 
A modified version 
of the classical Kalman filtering algorithm (referred to as KF-M) is also proposed that emerges from 
our EP derivations and is used in initializing the EP-based algorithm. Performance of Kalman 
Smoothing algorithm followed by KF-M (referred to as KS-M) is also examined.
Simulation results show that the 
performance of KF-M, KS-M, and EP algorithms improves 
with the increase in the number of base station antennas $M$ and the length of the data 
symbols $T_d$ in the transmitted frame. Thus for a fixed $M$, using a large $T_d$ 
in the semi-blind approach can mitigate the effect of pilot contamination in multi-cell systems.
Moreover, the EP-based algorithm significantly outperforming KF-M and KS-M algorithms.
Finally,
our results show that when applied to time-varying channels, 
these algorithms outperform 
the algorithms that are developed for block-fading channel models.

\bibliographystyle{IEEEtran}
\bibliography{References}



\end{document}